\documentclass[]{spie}  

 \usepackage[font=small,skip=0pt]{caption}
 \usepackage[square,comma,sort,numbers]{natbib}
\usepackage{amsmath,amsfonts,amssymb}
\usepackage{graphicx}
\usepackage{floatrow}
\usepackage{comment}
\usepackage{gensymb}
\usepackage{float}
\usepackage[section]{placeins}
\floatstyle{plaintop}
\restylefloat{table}
\newfloatcommand{capbtabbox}{table}[][\FBwidth]
\usepackage{blindtext}
\usepackage[colorlinks=true, allcolors=blue]{hyperref}

\title{Status of commissioning stabilized infrared Fizeau interferometry with LBTI}

\author[a,b]{Eckhart Spalding}
\author[c,d]{Phil Hinz}
\author[a,b]{Katie Morzinski}
\author[e,a]{Steve Ertel}
\author[a,b]{Paul Grenz}
\author[a,b]{Erin Maier}
\author[a,b]{Jordan Stone}
\author[e]{Amali Vaz}
\affil[a]{Steward Observatory, University of Arizona, Tucson, AZ 85721}
\affil[b]{Center for Astronomical Adaptive Optics, University of Arizona, Tucson, AZ 85721}
\affil[c]{Department of Astronomy \& Astrophysics, University of California Santa Cruz, Santa Cruz, CA 95064}
\affil[d]{Laboratory for Adaptive Optics, Center for Adaptive Optics, Santa Cruz, CA 95064}
\affil[e]{Large Binocular Telescope Observatory, Tucson, AZ 85721}

\authorinfo{Further author information: (Send correspondence to E.S.)\\E.S.: E-mail: spalding@email.arizona.edu}

\begin{document} 
\maketitle

\begin{abstract}
The Large Binocular Telescope Interferometer (LBTI) has the longest baseline in the world---22.7 m---for performing astronomical interferometry in Fizeau mode, which involves beam combination in a focal plane and preserves a wide field-of-view. LBTI can operate in this mode at wavelengths of 1.2--5 and 8--12 $\mu$m, making it a unique platform for carrying out high-resolution imaging of circumstellar disks, evolved stars, solar system objects, and possibly searches for planets, in the thermal infrared.

Over the past five years, LBTI has carried out a considerable number of interferometric observations by combining the beams near a pupil plane to carry out nulling interferometry. This mode is useful for measuring small luminosity level offsets, such as those of exozodiacal dust disks. The Fizeau mode, by contrast, is more useful for generating an image of the target because it has more $(u,v)$ (Fourier) plane coverage. 

However, the Fizeau mode is still in an ongoing process of commissioning. Sensitive Fizeau observations require active phase control, increased automation, and the removal of non-common-path aberrations (NCPA) between the science and phase beams. This increased level of control will increase the fringe contrast, enable longer integrations, and reduce time overheads.

We are in the process of writing a correction loop to remove NCPA, and have carried out tests on old and synthetic data. We have also carried out on-sky Fizeau engineering tests in fall 2018 and spring 2019. In this article, we share lessons learned and strategies developed as a result of these tests. 

\end{abstract}


\section{INTRODUCTION}
\label{sec:intro}  

The Large Binocular Telescope (LBT), located on Mt. Graham, Arizona, USA, is a stepping stone to next-generation extremely large telescopes (ELTs). The LBT is equipped with two telescopes on a single mount, both of which have 8.4 m primary mirrors and adaptive secondary mirrors to remove wavefront aberrations induced by the atmosphere. Both telescopes can be used simultaneously as separate unit telescopes (as long as they stay within co-pointing limits), or they can be used together to perform aperture synthesis and thus obtain resolutions at baselines that reach from one primary mirror to the other. Interferometry with the LBT allows this facility to reach resolutions and sensitivities approaching those of the multisegmented ELTs of the future.\footnote{See Appendix \ref{sec:appendix_glossary} for a quick-reference of the various acronyms and other terms used in this article.}

\begin{figure} [ht]
   \begin{center}
   \begin{tabular}{c} 
   \includegraphics[trim={4cm, 1.5cm, 3.0cm, 1.0cm}, clip=True, width=0.23\linewidth]{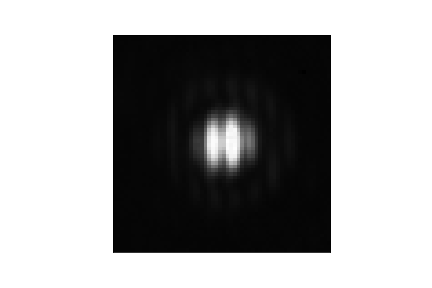}
   \includegraphics[trim={4cm, 1.5cm, 3.0cm, 1.0cm}, clip=True, width=0.23\linewidth]{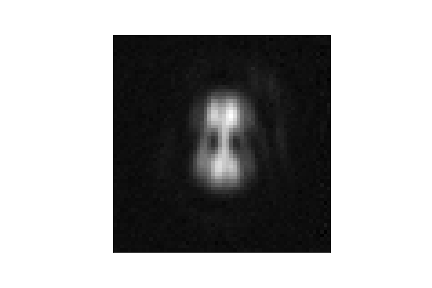}
   \includegraphics[trim={4cm, 1.5cm, 3.0cm, 1.0cm}, clip=True, width=0.23\linewidth]{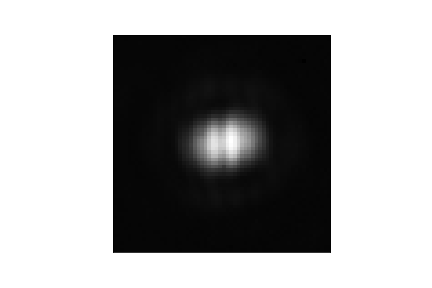}
   \includegraphics[trim={4cm, 1.5cm, 3.0cm, 1.0cm}, clip=True, width=0.23\linewidth]{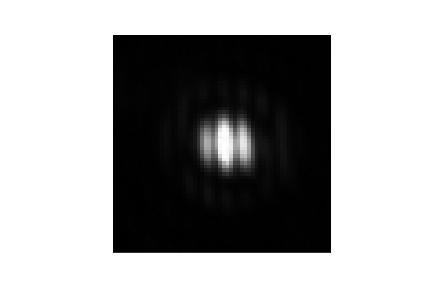}
   \end{tabular}
   \end{center}
   \vspace{-0.6cm}
   \caption[example] 
   { \label{fig:aberr_exmaples} 
Empirical full-aperture Fizeau illuminations on LMIRcam exhibiting differential aberrations. Left to right: differential OPD; differential tip; differential tilt; and a near-perfect illumination, though the optical path location within the coherence envelope is unknown. Greyscale is linear. Each side of the subplots is $\approx 3/4$ asec. }
\end{figure}

The first adaptive-optics-corrected interferometric fringes were obtained with the LBT in 2012 \cite{hinz2012first}. Since then, the Large Binocular Telescope Interferometer (LBTI) has performed most science interferometry in ``nulling'' mode as part of the HOSTS survey \cite{ertel2018hosts} to detect exozodiacal dust disks at sensitivities down to the order to tens of solar system zodiacal disks. A second interferometric mode, ``Fizeau'', involves beam overlap in the focal plane and paints out $(u,v)$-space out to the 22.7-m edge-to-edge mirror separation. 

In \cite{spalding2018towards} we described some of the remaining obstacles to commissioning the Fizeau mode. Firstly, the science and phase optical trains have a greater non-common-path configuration than in nulling mode, and greater care has to be taken to co-align each pair of beams. In nulling mode, the beam combination happens once, upstream of the phase and the NOMIC science camera. (Nulling is currently not possible with LMIRcam.) The phase-sensing camera is blind to the illumination pattern on the science detectors in both nulling and Fizeau mode, but in the Fizeau mode the problem is worse because the science and phase beam pairs have to interfere separately. (See Appendix \ref{sec:appendix_align} and illustrations in \cite{spalding2018towards}.)

The Fizeau-mode illumination on the science detector also has more degrees of freedom than in nulling mode. In nulling mode, the science detector illumination is (to first order) an Airy function. In Fizeau mode, the illumination exhibits fringes from the interference of the two beams on the detector itself. But even if the plane of the phase detector is at the center of the coherence envelope and the phase loop is closed, the illumination on the science detector can exhibit differential tip, tilt, or optical path difference (Fig.\ \ref{fig:aberr_exmaples} and Appendix \ref{sec:appendix_align}). If the plane of the science detector is indeed outside of the coherence envelope, no fringes will be visible, and the PSF converges to an Airy function of incoherently-overlapped beams.

To counteract the decoupling between the phase and science channels in Fizeau mode, we are developing a correction code that automates as much of the alignment process as possible, and uses the science detector illumination in realtime to provide corrective movements to internal mirrors and to setpoints of the phase-tracking proportional–-integral–-derivative (PID) control loop. 

In this article, we briefly describe the currently-available Fizeau modes in Sec.  \ref{sec:currently_available}, changes to the telescope and instrument in the past year in Sec.  \ref{sec:expanding}, the alignment and correction software development in Sec.  \ref{sec:corr_code}, and lessons learned from on-sky tests in Sec.  \ref{sec:lessons}. We mention future steps and conclude in Sec.  \ref{sec:future}.

\section{Currently available Fizeau modes}
\label{sec:currently_available}

\begin{figure} [ht]
   \begin{center}
   \begin{tabular}{c} 
   \includegraphics[width=0.9\linewidth]{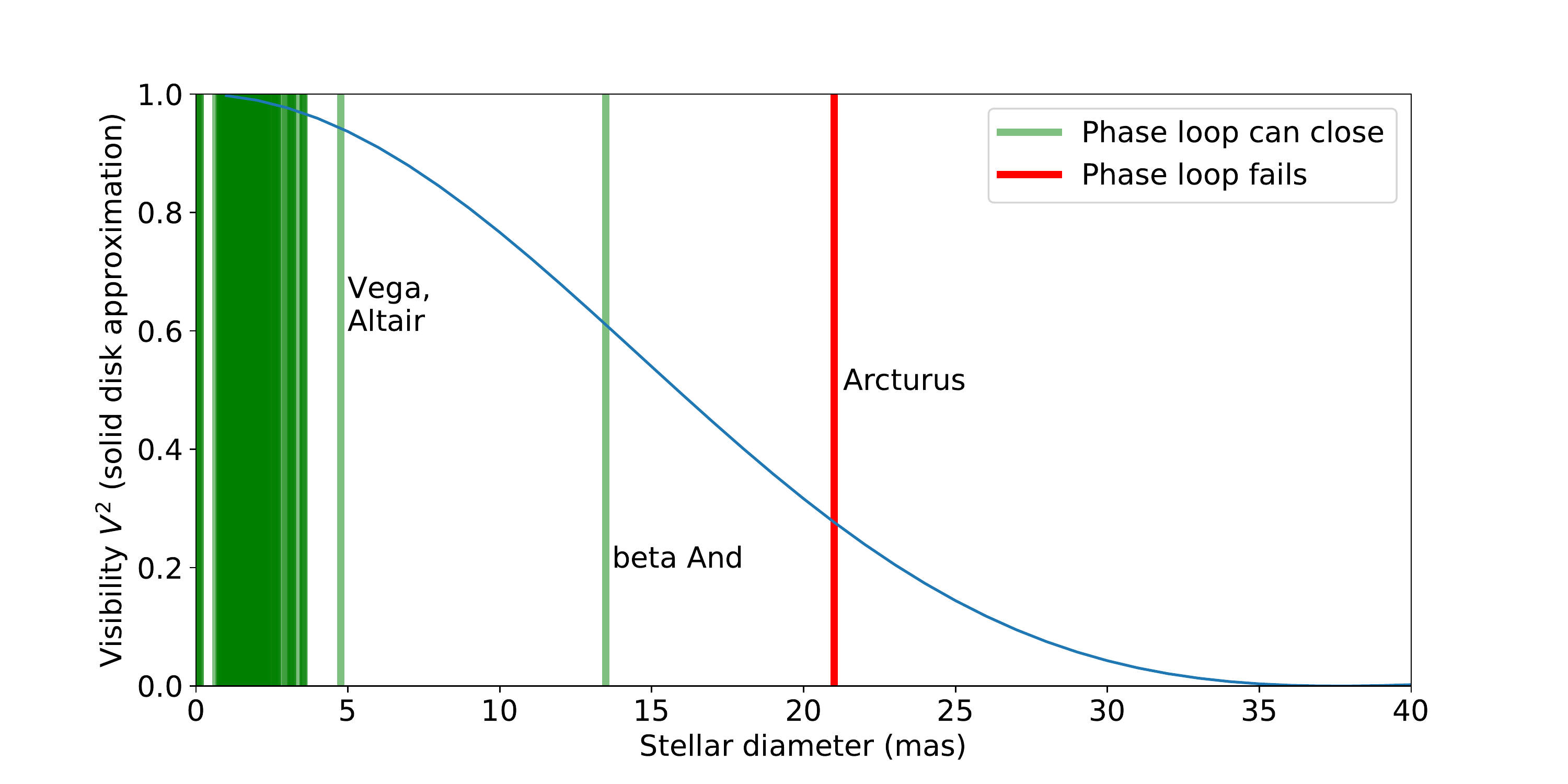}
   \end{tabular}
   \end{center}
   \vspace{-0.7cm}
   \caption[example] 
   { \label{fig:PhaseCam_vis_limits} 
PhaseCam visibility limitations. The visibility curve corresponds to that expected for a solid disk in the $Ks$-band (2.15 $\mu$m). Green lines are set down at stellar widths based on the $K$-band uniform disk approximation for all 137 stars on which PhaseCam has ever successfully closed, and which also appear in the JMMC Stellar Diameter Catalog \cite{chelli2016pseudomagnitudes}. (These stellar diameters are calculated with polynomial fits and magnitudes in two different bands, though some of these stars have also had their diameters measured directly with interferometry.) Most of these stars were observed as science or calibrator targets for the HOSTS survey \cite{ertel2018hosts}. The red line is set down at the $H$-band width of Arcturus \cite{lacour2008limb}, on which PhaseCam fails to close.}
\end{figure}

\noindent
The number of targets observed with LBTI's Fizeau mode remain very few in number. Table \ref{table:current_targets} shows the targets which have appeared in either conference proceedings articles or the peer-reviewed literature. All of those observations were made without active phase control, which requires very bright and point-like targets (see Table \ref{table:observing_reqs} and Fig.\ \ref{fig:PhaseCam_vis_limits}). There exist additional science Fizeau datasets, including one with partial phase control, which are currently undergoing reduction. Here we describe the currently-available Fizeau modes.

\begin{table}[!htbp]
\begin{center}
\caption{LBTI Fizeau targets in the literature} 
\label{table:current_targets}
\vspace{-0.3cm}
\begin{tabular}{| c | c | c | l | c |}
\hline
\textit{Target}	& \textit{Mode} & 	\textit{Wavelength}	& \textit{Remarks}  & \textit{Ref} \\
 \hline	
CH Cyg + calib	& Fizeau-Airy & 4 $\mu$m & \begin{tabular}{@{}l@{}}Test target; decrease in fringe\\visibility appears in CH Cyg\\because of stellar outflows.\end{tabular} & \cite{hinz2012first,hill2013large} \\
\hline
Trapezium asterism & Fizeau-Airy & 4 $\mu$m 	&\begin{tabular}{@{}l@{}}Test target; demonstration of\\co-phasing across $\approx$7 arcsec\end{tabular} 	&  \cite{hinz2014commissioning} \\	
\hline
Vega & Fizeau-Airy &	11 $\mu$m	& Test target	& \cite{hoffmann2014operation}  \\
\hline
LkCa 15 + calibs & 	NRM & 2.2 and 3.7 $\mu$m	&\begin{tabular}{@{}l@{}}Used baselines contained within\\each 8.4-m primary\end{tabular}	& \cite{sallum2015accreting,Sallum_2017} \\
\hline
MWC 349A + calib & NRM & 3.8 $\mu$m	& \begin{tabular}{@{}l@{}}Used baselines across the\\23-m dual aperture\end{tabular}	& \cite{sallum2017improved,Sallum_2017} \\
\hline
Io + calibs & Fizeau-Airy & 4.8 $\mu$m	& \begin{tabular}{@{}l@{}}First science target in full-aperture\\Fizeau mode; ``lucky'' fringing\end{tabular}	& \cite{leisenring2014fizeau,conrad2015spatially,conrad2016role,de2017multi} \\
\hline
\end{tabular}
\noindent
\end{center}
\end{table}

\begin{table}
\begin{center}
\caption{Current observing target requirements for Fizeau observations (updated from \cite{spalding2018towards}).} 
\label{table:observing_reqs}
\vspace{-0.3cm}
\begin{tabular}{| l  | l | l |}
\hline
\textit{Parameter} & \textit{Requirement}  & \textit{Remarks} \\
\hline
DEC   & $\gtrsim -5^{\circ}$ & \begin{tabular}{@{}l@{}l@{}}Constrained by the need for $\leq$1.2'' seeing \end{tabular} \\
 \hline
\begin{tabular}{@{}l@{}}$R$-band brightness \\ of AO guide star \end{tabular} & \begin{tabular}{@{}l@{}l@{}}$m_{R}\lesssim$12.5 mag (for 300 deformable\\modes, 40$\times$40 pupil subapertures,\\1 kHz)\end{tabular} & \begin{tabular}{@{}l@{}l@{}}The SOUL upgrade has been made to both\\the left and right telescopes. More precise\\limits remain to be determined. Note AO\\guide stars have been acquired as far as $\sim$30''\\off-axis from the science target\end{tabular} \\
\hline
\begin{tabular}{@{}l@{}l@{}}$K$-band brightness\\of phase star\end{tabular}& \begin{tabular}{@{}l@{}l@{}} $m_{K}\lesssim4.7$ for correction as slow\\as 520 Hz  \end{tabular}&     \begin{tabular}{@{}l@{}l@{}} For fringe tracking with PhaseCam at \\standard detector binning. In principle the\\phase star can be up to a few  arcseconds\\away from the science target. \end{tabular} \\
 \hline
\begin{tabular}{@{}l@{}l@{}}Visibility $V^{2}$\\of phase star\end{tabular}& \begin{tabular}{@{}l@{}l@{}}$V^{2}\gtrsim 0.6$ in $K_{S}$-band works;\\$0.6\gtrsim V^{2}\gtrsim 0.3$ is uncharacterized;\\$V^{2}\lesssim0.3$ fails\end{tabular}& \begin{tabular}{@{}l@{}l@{}} PhaseCam cannot lock onto extended\\ sources. (See Fig.\ \ref{fig:PhaseCam_vis_limits}.) \end{tabular} \\
 \hline
Science wavelength & 
\begin{tabular}{@{}l@{}l@{}} $L$-, $M$-, or $N$-bands \\  (limited sensitivity in $K$-band) \end{tabular} & \begin{tabular}{@{}l@{}l@{}}Limited $K$-band is possible by reflecting some\\of this into PhaseCam and some of it towards\\LMIRcam. \end{tabular} \\

\hline
\end{tabular}
\noindent
\end{center}
\end{table}

\subsection{Fizeau-Airy mode}

The classical LBTI Fizeau PSF is that produced using filled sub-apertures, without any additional wavelength dispersion. (See Fig.\ \ref{fig:aberr_exmaples}, or \ref{fig:airy_demos} or bottom-left panels in Fig.\ \ref{fig:tilt_spie}.) This PSF is a multiplication of an Airy function with a corrugation from the separation of the two sub-pupils. The filled apertures maximize the probed $(u,v)$ space, and as such it is best suited for reconstructing detailed images.

\subsection{Non-redundant phase masking (NRM)}

LBTI currently has two sets of pupil masks which are peppered with holes to provide non-redundant baselines across the pupil. One set of masks has a pattern of 12 holes, the other 24. At a steep cost of throughput, these baselines---either contained within a single telescope aperture or across both apertures---allow a fine characterization of the stellar PSF and its subsequent removal. With fast readouts, ``closure phase'' across triangles of baselines provides a form of phase control even in the absence of a mechanical phase control.

\subsection{Fizeau-grism mode (i.e., spectrointerferometry)}

This mode involves the dispersion of the Fizeau-Airy PSF with a grism. This effectively extends the coherence envelope by reducing the wavelength bandpass to the wavelengths received by each row of pixels perpendicular to the dispersion axis. This mode is useful for low-spectral-resolution spectroscopy of bright targets at high spatial resolution. This is particularly useful if the object is extended enough that the fringe visibility is too low in Fizeau-Airy mode. (See Fig.\ \ref{fig:grism_demos}.) \footnote{The Fizeau-Airy PSF can be thought of as an marginalization along the dispersion axis of the Fizeau-grism PSF. Fringes may have high contrast in the Fizeau-grism PSF, but fringes which are slightly displaced in each row of pixels can wash out after integrating over those rows of pixels.}

\section{Expanding the science capabilities}
\label{sec:expanding}

\subsection{Hardware changes to telescope}

AO correction is a prerequisite for sensitive infrared interferometry by pooling science photons into a high-Strehl PSF with minimal speckle noise, a frozen fringe pattern, and a minimal footprint on top of the high sky background. Up to one year ago, the two AO systems and the LBTI wavefront sensors could correct for atmospheric aberrations at up to 1.0 kHz on bright targets, using up to 30$\times$30 correction subapertures in the pupil. In the summer of 2018, the left-side telescope LBTI wavefront sensor was upgraded as part of the SOUL project with detectors with less read noise and faster readouts \cite{pinna2016soul,christou2018adaptive}.

The SOUL upgrade increased the maximum correction frequency from 1.0 to 1.7 kHz, the maximum number of subapertures from 30$\times$30 to 40$\times$40, and the maximum number of controlled mirror modes from 400 to 500. In January and February 2019, the right-side wavefront sensor was also upgraded with SOUL.  These upgrades will offer higher Strehl and greater tolerance of atmospheric conditions. In addition, work is ongoing to improve the vibration feed-forward system Optical Path Difference and Vibration Monitoring System Plus (OVMS+), so as to feed in better predicted changes in pathlength to the phase-sensing PID loop \cite{bohm2016ovms,bohm2017improving}. (See Sec.  \ref{subsec:phase_ctrl}.) This will reduce the phase noise in closed phase loop while doing interferometry, and reduce the probability that the phase loop will break entirely.

\begin{figure} [ht]
   \begin{center}
   \begin{tabular}{c} 
   \includegraphics[width=0.9\linewidth]{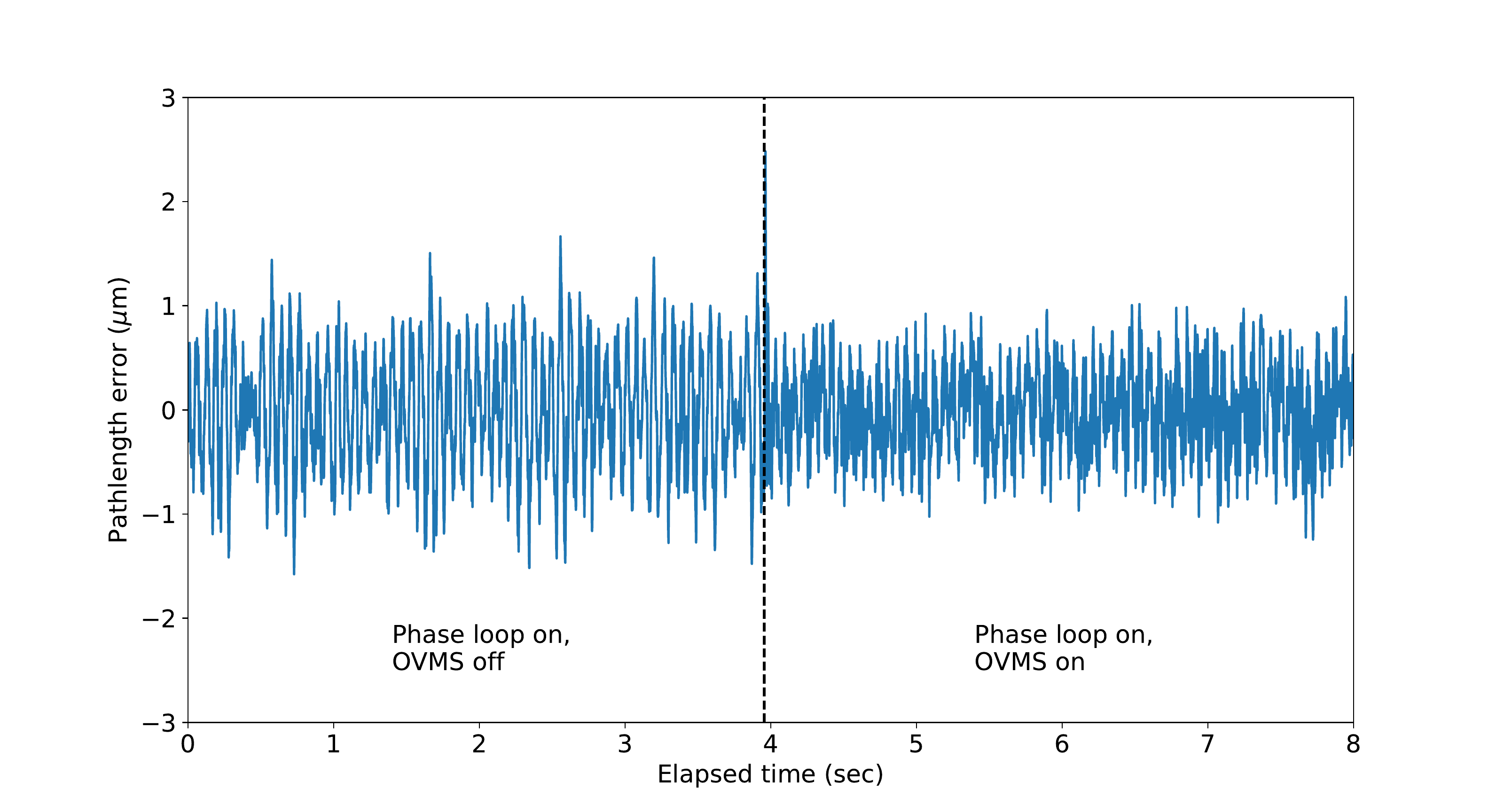}
   \end{tabular}
   \end{center}
   \vspace{-0.7cm}
   \caption[example] 
   { \label{fig:ovms_demo} 
Improvement from the OVMS feed-forward to the phase loop, from UT 2019 April 20.}
\end{figure}

The most recent comparison of the quality of the OVMS feed-forward to the phase loop was on UT 2019 April 20, where pathlength changes were 0.57 $\mu$m rms with the phase loop closed but without OVMS, and 0.43 $\mu$m rms with the phase loop closed and OVMS on (Fig.\ \ref{fig:ovms_demo}). During a phase-controlled Fizeau observation on UT 2018 May 7, path length rms was as low as 0.30 $\mu$m.

\subsection{Phase control}
\label{subsec:phase_ctrl}

\begin{figure} [ht]
   \begin{center}
   \begin{tabular}{c} 
   \includegraphics[height=5.6cm, trim={1.5cm, 0.5cm, 2cm, 0.5cm}, clip=True]{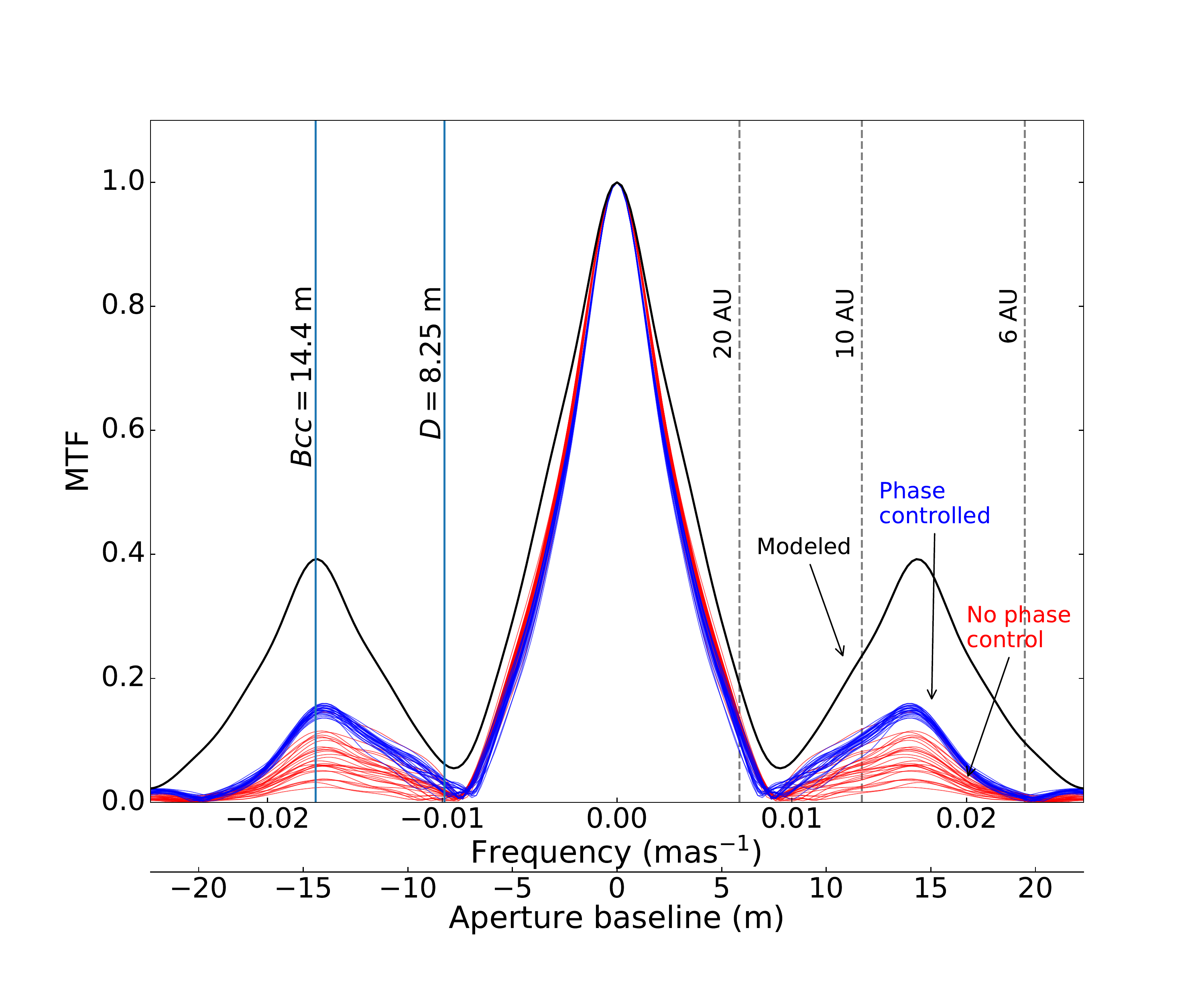}
   \includegraphics[height=5.6cm, trim={1.5cm, 0cm, 2cm, 0cm}, clip=True]{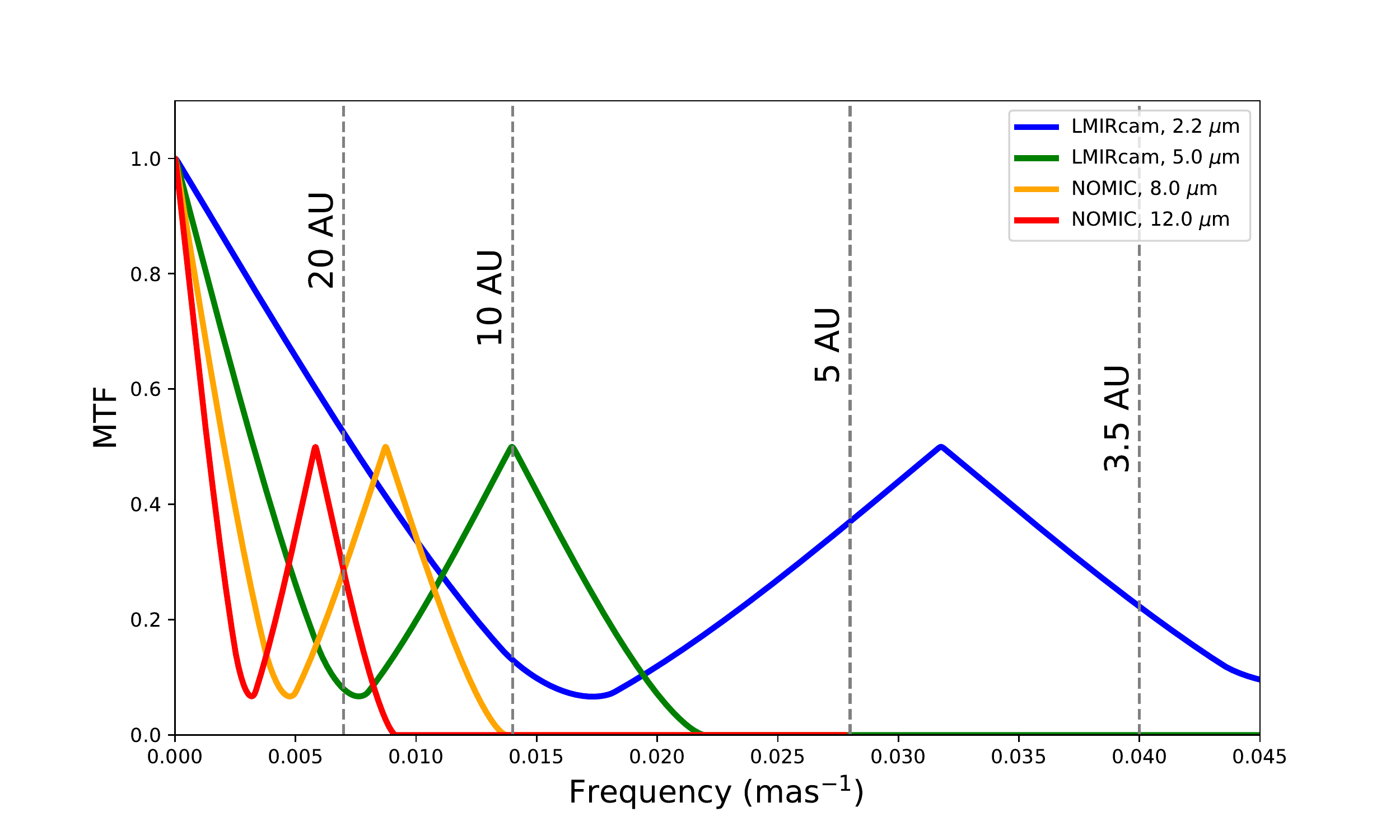}
   \end{tabular}
   \end{center}
   \vspace{-0.6cm}
   \caption[example] 
   { \label{fig:cross_mtf} 
Left: MTFs for LBTI's 4.01-4.08 $\mu$m filter. Black is generated using a simulated polychromatic PSF. Blue is a sampling of empirical MTFs when phase control was active. Red is the same number of samples without phase control. Vertical lines show spatial scales in AU for Taurus-Auriga, the nearest large star-forming complex at 140 pc, and in meters the equivalent baselines of a stopped-down primary mirror and the center-to-center mirror baseline. The decrease in amplitude of the high-frequency lobes of the empirical MTFs is consistent with predictions of \cite{patru2017lbti} for MTFs in the presence of AO residuals, differential piston errors, or phase smearing during an integration. Furthermore, these MTFs are polychromatic and have finite coherence envelope lengths. A nonzero OPD from the center of the coherence envelope will also decrease the PSF fringe contrast and the amplitude of the high-frequency lobes of the MTF. Right: A cross-section of the MTF along the long baseline, for different wavelengths accessible to the LBTI's science cameras LMIRcam (1.2--5 $\mu$m) and NOMIC (8--12 $\mu$m). Vertical lines again indicate spatial scales corresponding to Taurus-Auriga. }
\end{figure}

\begin{figure} [ht]
   \begin{center}
   \begin{tabular}{c} 
   \includegraphics[width=0.8\linewidth, trim={0cm, 23cm, 0cm, 0cm}, clip=True]{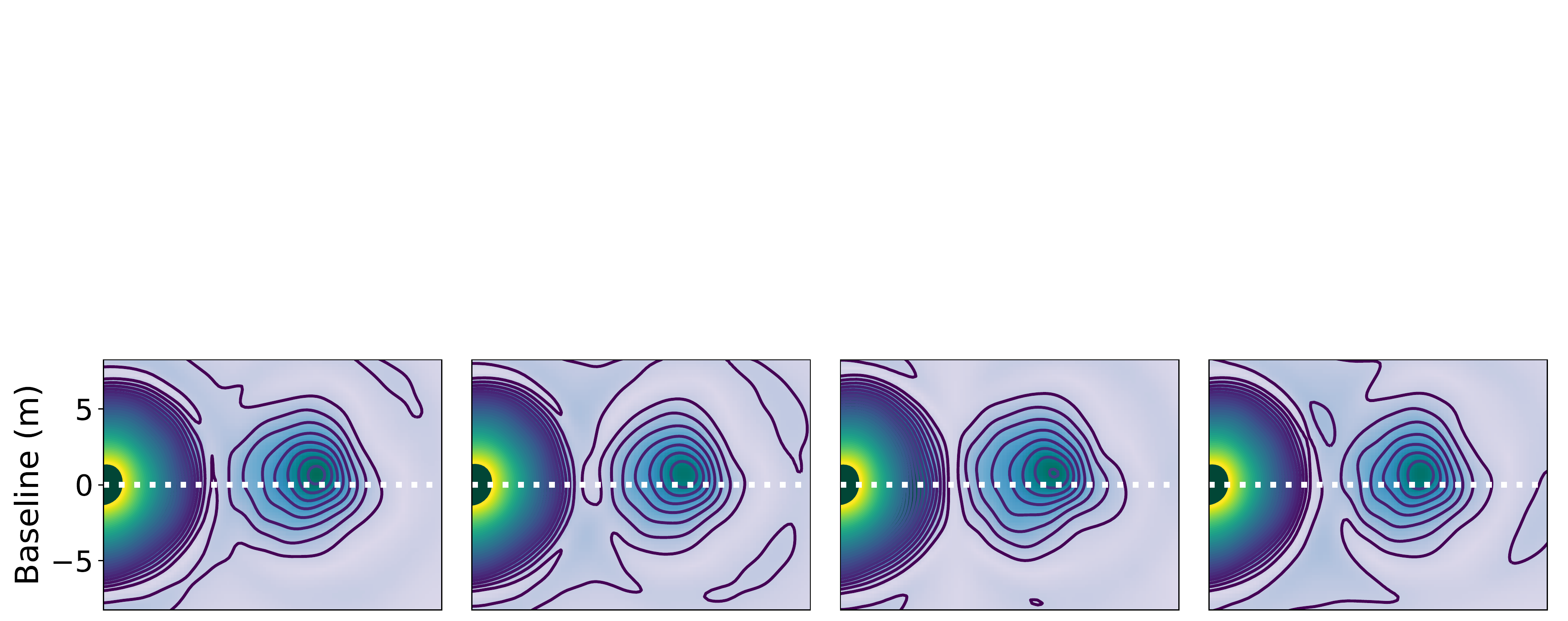} \\
   \includegraphics[width=0.8\linewidth, trim={0cm, 7cm, 0cm, 23cm}, clip=True]{images/contour_mtfs.pdf} \\
   \includegraphics[width=0.8\linewidth, trim={0cm, 0cm, 0cm, 28cm}, clip=True]{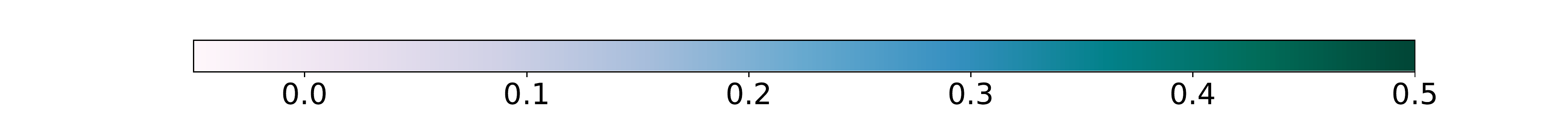}
   \end{tabular}
   \end{center}
   \vspace{-0.6cm}
   \caption[example] 
   { \label{fig:contours} 
Top row: Contour plots of the MTFs of eight different PSFs, showing how phase control puts more power into the high-frequency lobes. All colors and contours are linear, and are set to those of the MTF in the top left. Top row: MTFs with phase control. Bottom row: MTFs without phase control. White dotted lines are for reference. There appears a slight asymmetry along the short baseline of the location of the high-frequency node peak (i.e., there is a slight slant to the phase-controlled fringes on the detector; see rightmost panel in Fig.\ \ref{fig:aberr_exmaples}). This asymmetry along the short baseline is not evident in the MTFs without phase control.}
\end{figure}

Phase control coupled with the Fizeau mode improves the sampling of Fourier space at high frequencies \cite{patru2017lbti}. This sampling of Fourier space can be quantified with the complex optical transfer function (OTF), which has an amplitude (the modulation transfer function, or MTF) and a phase (the phase transfer function, or PTF). The MTF of the LBT in Fizeau mode has a characteristic triple-peaked shape along the long baseline, and which stretches out to frequencies equivalent to the edge-to-edge mirror separation. (See Figs. \ref{fig:cross_mtf}, \ref{fig:contours}.)

LBTI has been controlling the OPD between the two telescope beams since 2013 with the Phasecam camera, which is based on a PICNIC detector \cite{defrere2014co}.  However, the detector was installed in anticipation of correcting the phase on very bright, unresolved stars in the HOSTS target list \cite{weinberger2015target,jordan_fftcam}. As such, the read noise of this detector limits phase-controlled targets to $Ks\lesssim4.7$. Fringe visibility also decreases for more extended targets due to the angular diversity of the wavefronts, to the point where PhaseCam cannot lock onto an object if it is extended, even if it is bright enough. In Fig.\ \ref{fig:PhaseCam_vis_limits} we show the visibility limits of PhaseCam. 

The PhaseCam PID software remains in the same state as it was at the completion of the HOSTS survey in 2018. The phase loop was closed for the first time during a Fizeau science observation in May 2018, and Fig.\ \ref{fig:contours} shows examples of the MTF with and without phase control. Though there are no immediate plans for modifying the PID loop itself, we are supplementing the PID loop with software which uses the PhaseCam $H$-band illumination to automatically correct $Ks$-band phase ``jumps'' \cite{maier2018two}, which occur when an atmospherically-induced  phase shift happens quickly enough that the PID loop latches on to the wrong fringe. (Until now, corrections have required manual intervention.) 

It should be noted that science can be done \textit{without} phase control, albeit at reduced sensitivity. The Fizeau correction code described below will also be able to partly compensate for an open phase loop by analyzing the science detector illuminations and making periodic pathlength corrections.

\begin{figure} [ht]
   \begin{center}
   \begin{tabular}{c} 
   \includegraphics[trim={0.5cm, 1.8cm, 7.5cm, 2.2cm}, clip=True, width=0.7\linewidth]{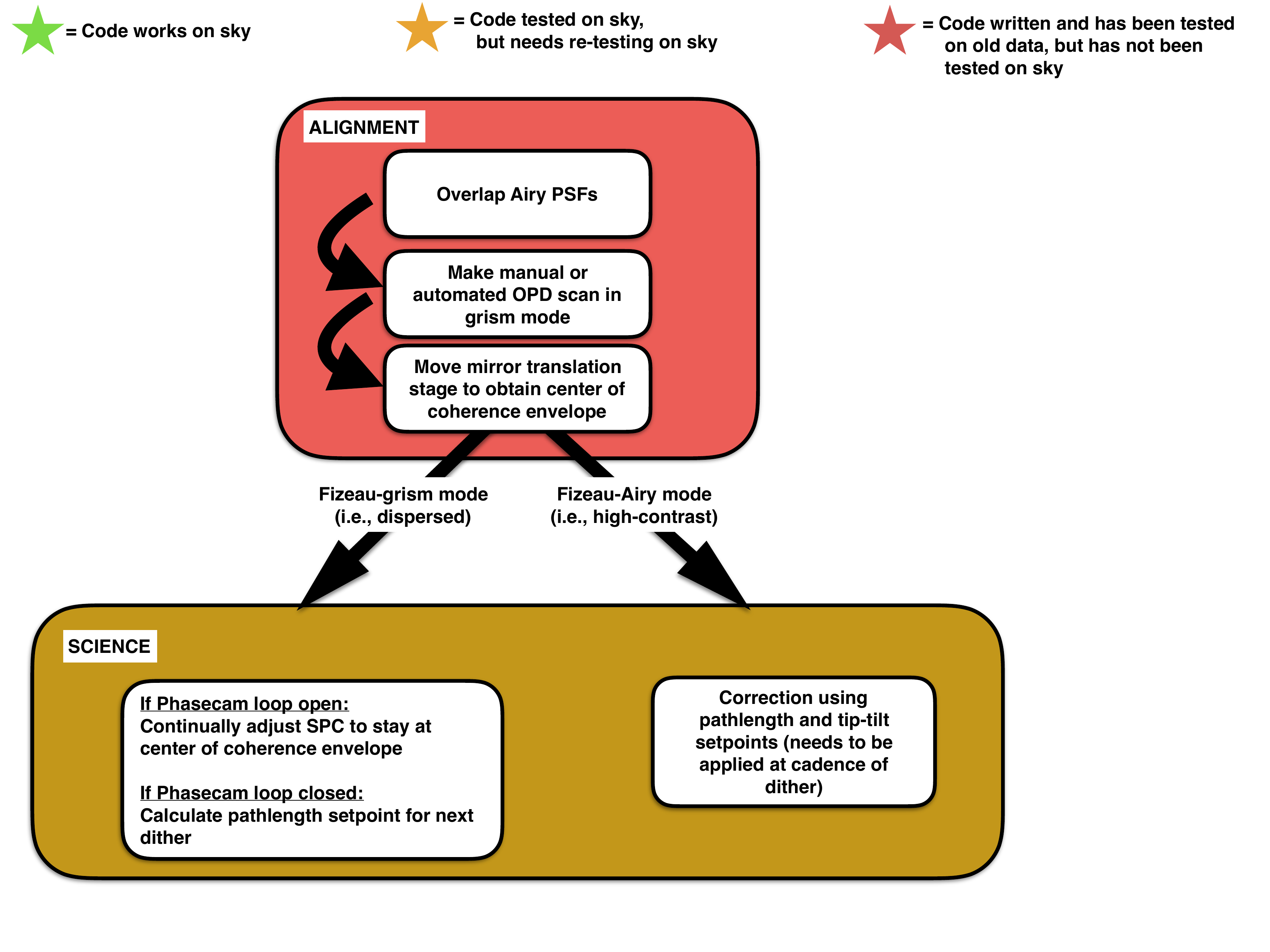}
   \end{tabular}
   \end{center}
   \vspace{-0.6cm}
   \caption[example] 
   { \label{fig:code_status} 
The sequence of steps which the Fizeau alignment and correction loop will automate. The parts of this code are at various stages of testing and implementation. For more description of the alignment steps, see Sec.  \ref{subsec:initial_align}; for steps at the science stage, see Sec.  \ref{subsec:corrxn_code}.}
\end{figure}

\subsection{Fizeau alignment/correction code development}

Over the past year we have been writing software to make alignments immediately prior to Fizeau observations, and to run a correction loop to remove differential aberrations on the science detector during observations (Fig.\ \ref{fig:code_status}). The pathfinding version of this code is being written in Python, together with INDI \cite{indiweb} telescope and instrument control commands. See Sec.  \ref{sec:corr_code} for more details.\\

\vspace{-3cm}

\section{Fizeau alignment/correction code}
\label{sec:corr_code}

\subsection{Initial alignment}
\label{subsec:initial_align}

After the AO loops are closed, a script overlaps the Airy PSFs on the detector by sending small movement commands to the telescopes. Next, the OPD between both beams is brought to zero, or at least as close to the middle of the coherence envelope as can be determined. This is done by dispersing the illuminations with a grism. This expands the coherence envelope along each row of pixels in ($x,y$)-space, and the angle of the fringes can be found by taking a Fourier transform and localizing the corresponding `bump' in the 2D MTF of the Fourier transform, in ($\zeta,\eta$)-space.

\begin{figure} [ht]
   \begin{center}
   \begin{tabular}{c} 
   \includegraphics[width=0.3\linewidth, trim={0cm 0cm 0cm 0cm}, clip=True]{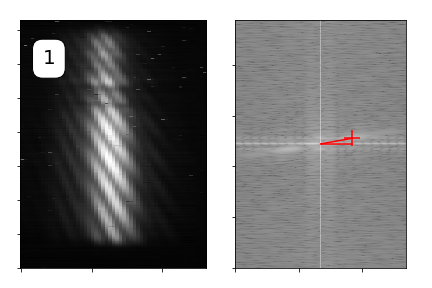}
   \vline
   \includegraphics[width=0.3\linewidth, trim={0cm 0cm 0cm 0cm}, clip=True]{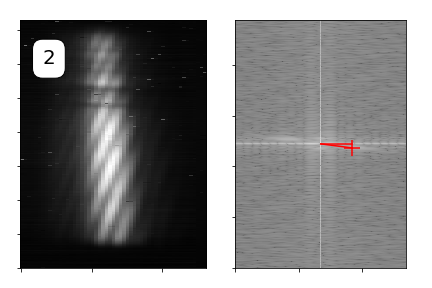}
   \vline
   \includegraphics[width=0.3\linewidth, trim={0cm 0cm 0cm 0cm}, clip=True]{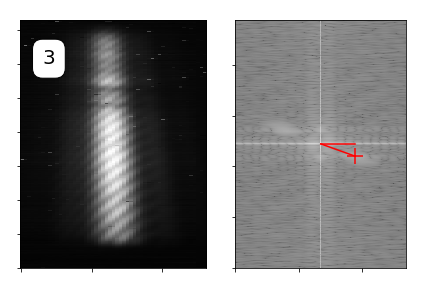}
   \end{tabular}
   \end{center}
   \vspace{-0.6cm}
   \caption[example] 
   { \label{fig:grism_demos} 
Empirical Fizeau-grism illuminations on LMIRcam at different OPD (stretched in x for display) and their MTFs (in logarithmic greyscale). The angle in red is determined by finding the bump in Fourier space corresponding to the frequency content of the fringes. Numbers correspond to those in the left-hand plot of Fig.\ \ref{fig:greystuff}.}
\end{figure}

If the dispersion axis of the grism is parallel to the $y$-axis, the OPD is approximately proportional to the tangent of the angle between the $+x$-axis and a line normal to the fringes, or equivalently, of the bump in Fourier space with the $+\zeta$-axis. (See Fig.\ \ref{fig:grism_demos}.) The HPC mirror is shifted along a translation stage until the pathlength causes the fringes to be parallel to the grism dispersion axis on the detector (or equivalently, the angle of the bump in Fourier space is brought to zero). Since the wavelengths vary along the grism illumination, there will be some dispersion of the power in Fourier space. This effect is not important at this stage, as long as the fringes can be made as parallel with the grism dispersion axis as possible.
\footnote{For any intermediary wavelength $\lambda_{S}<\lambda<\lambda_{L}$ between the shortest and longest wavelengths $\lambda_{S}$ and $\lambda_{L}$, the angles $\alpha$ with which fringes at those wavelengths form with the dispersion axis of the grism on the detector are $\frac{tan[\alpha(\lambda_{S})]}{tan[\alpha(\lambda)]} = \frac{\lambda}{\lambda_{S}}$. In the small angle limit, the variation in these angles along the bandpass is $\alpha(\lambda_{S})/\alpha(\lambda) \approx \lambda/\lambda_{S}$. The most atmospherically transmissive region of the 2.8-4.2 $\mu$m grism is roughly 3.3-4.2 $\mu$m, for which $\alpha(\lambda_{S})/\alpha(\lambda_{L}) \approx 1.27$. An alternative strategy is to make all the fringes from the detector parallel by remapping the coordinates from OPD and wavelength to phase and wavenumber: $\tau\rightarrow\phi$, $\lambda\rightarrow\kappa$  (e.g., \cite{improvements,practical}).}

\begin{figure} [ht]
   \begin{center}
   \begin{tabular}{c} 
   \includegraphics[width=0.3\linewidth, trim={0cm 1.5cm 0cm 1cm}, clip=True]{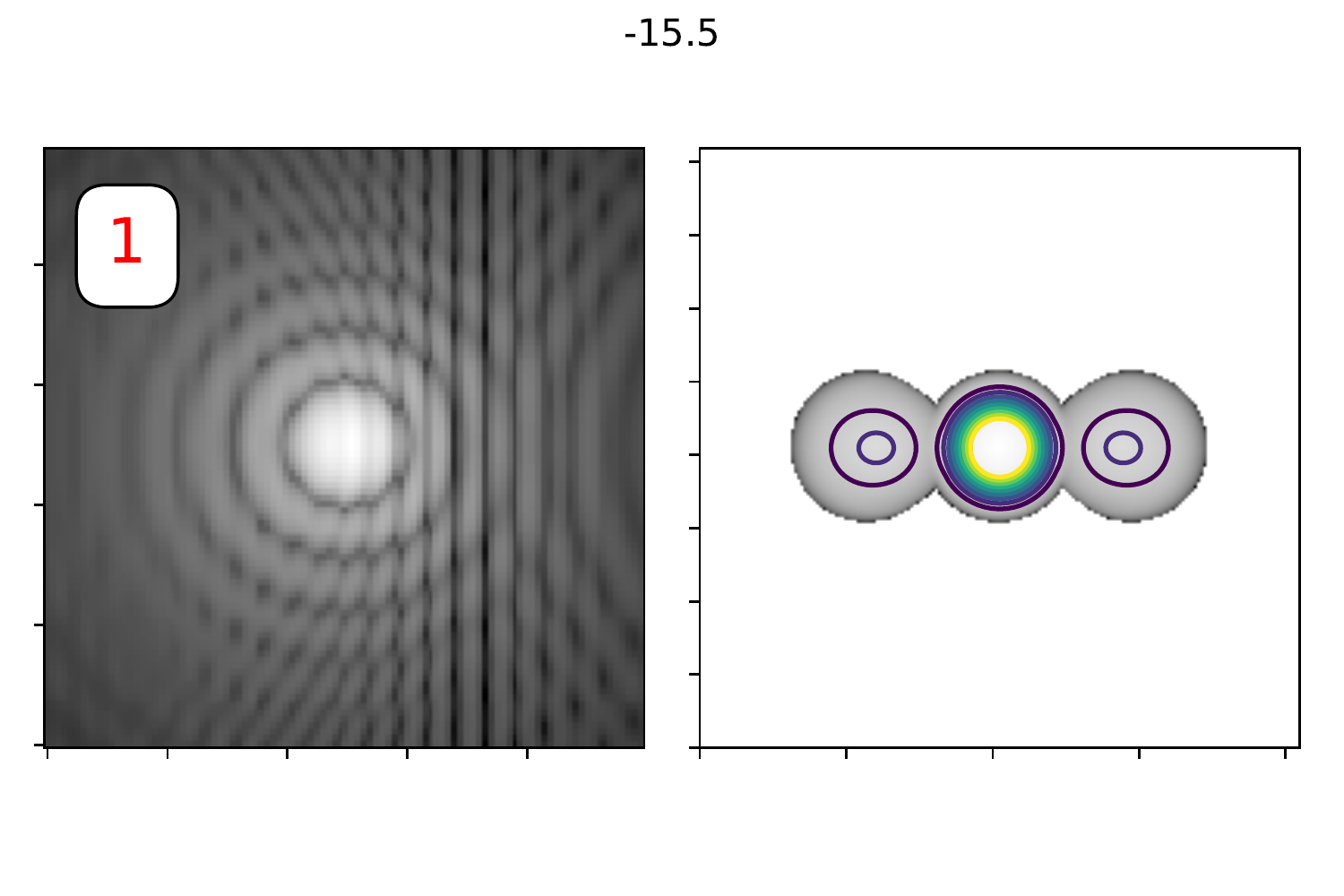}
   \vline
   \includegraphics[width=0.3\linewidth, trim={0cm 1.5cm 0cm 1cm}, clip=True]{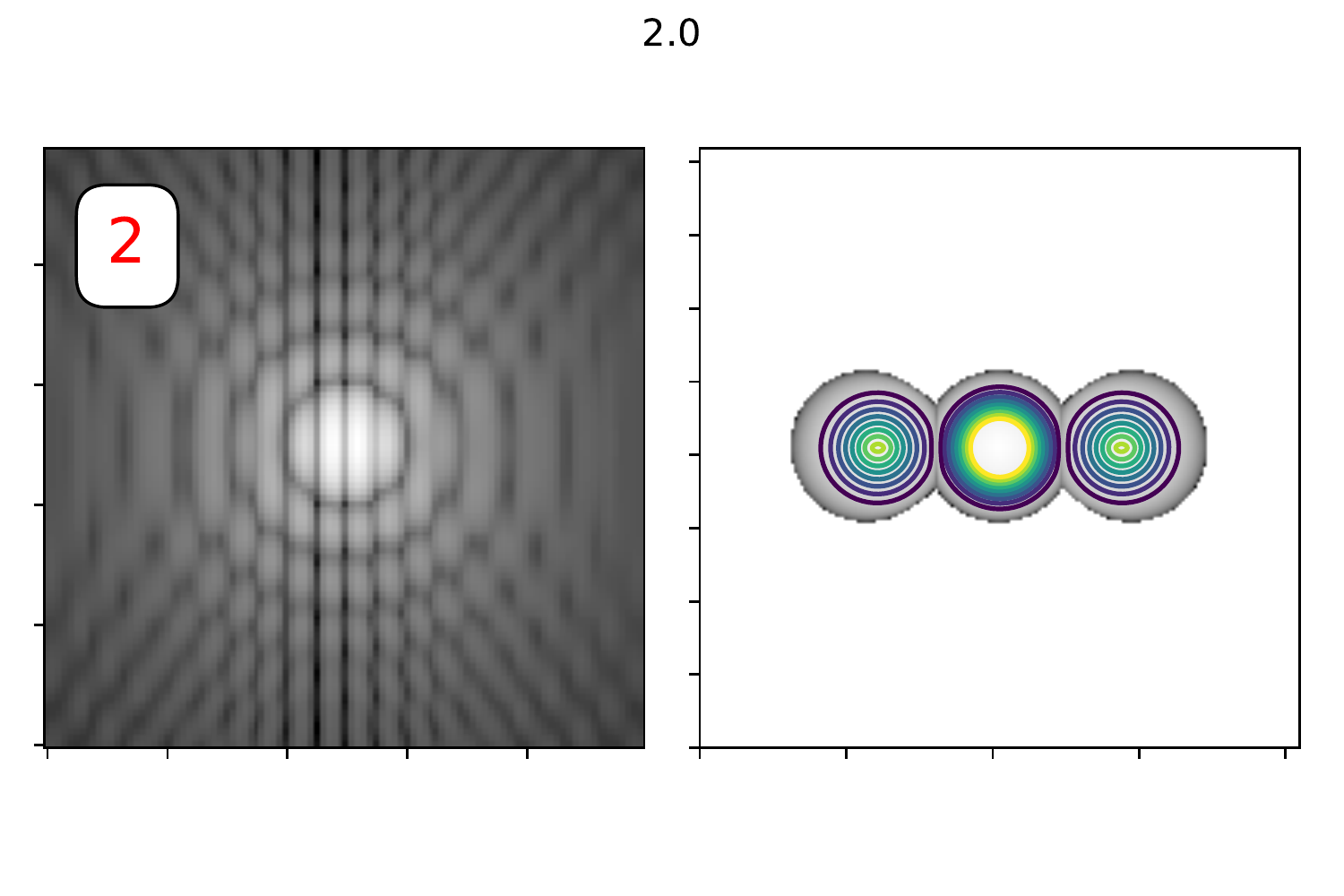}
   \vline
   \includegraphics[width=0.3\linewidth, trim={0cm 1.5cm 0cm 1cm}, clip=True]{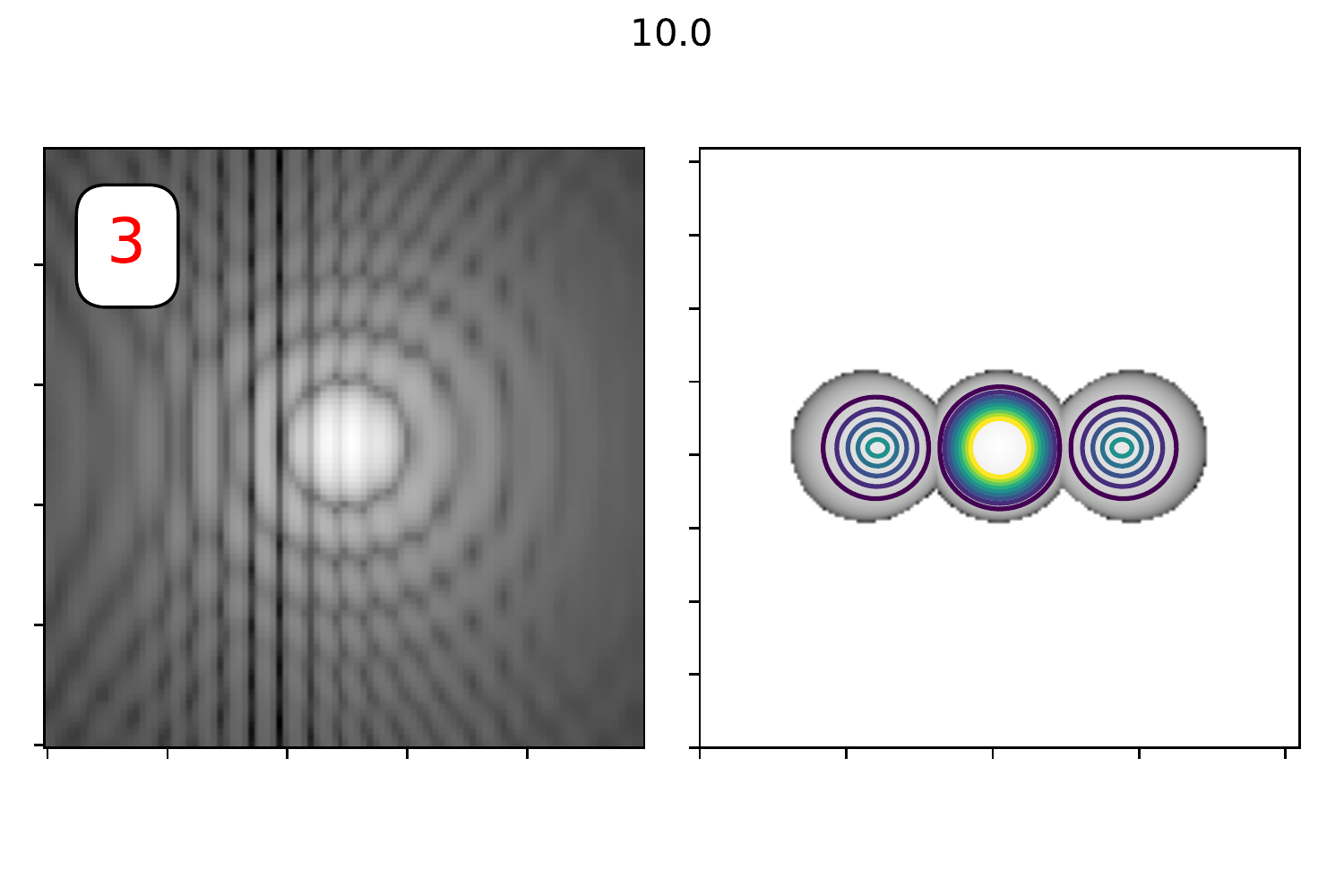}
   \end{tabular}
   \end{center}
   \vspace{-0.6cm}
   \caption[example] 
   { \label{fig:airy_demos} 
Simulated Fizeau-Airy illuminations at different OPD and their MTFs (both in logarithmic greyscale; the MTF contours are all on the same color scale). The middle lobe of the MTFs corresponds to spatial information from baselines within each 8.25-m aperture. Off-center lobes encode the high frequencies from baselines stretching across both LBT sub-apertures. Numbers correspond to those in the right-hand plot of Figs. \ref{fig:greystuff}.}
\end{figure}

\begin{figure} [ht]
   \begin{center}
   \begin{tabular}{c} 
   \includegraphics[height=5.5cm, trim={0cm 0cm 0cm 0cm}, clip=True]{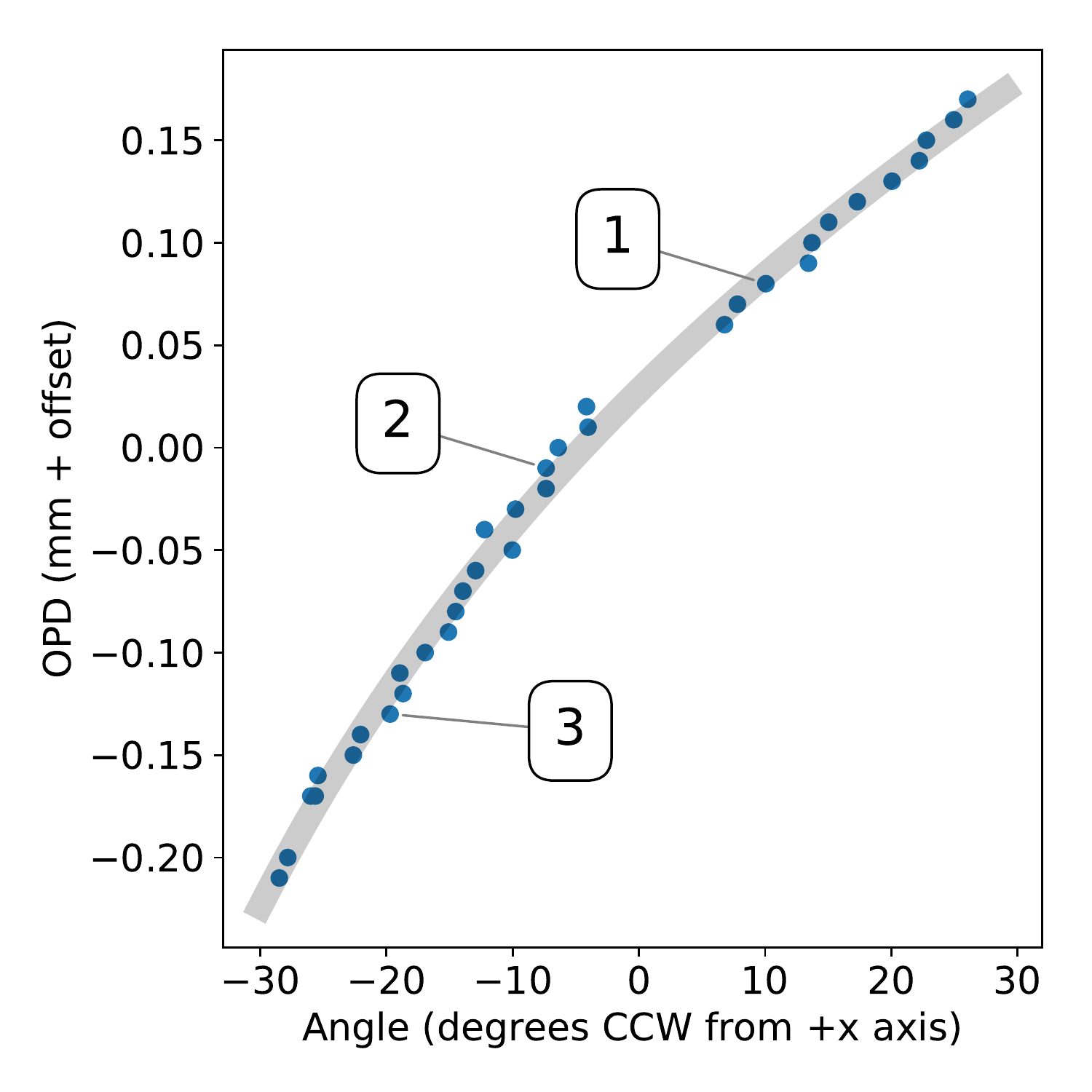}
   \includegraphics[height=5.5cm, trim={0cm 0cm 0cm 0cm}, clip=True]{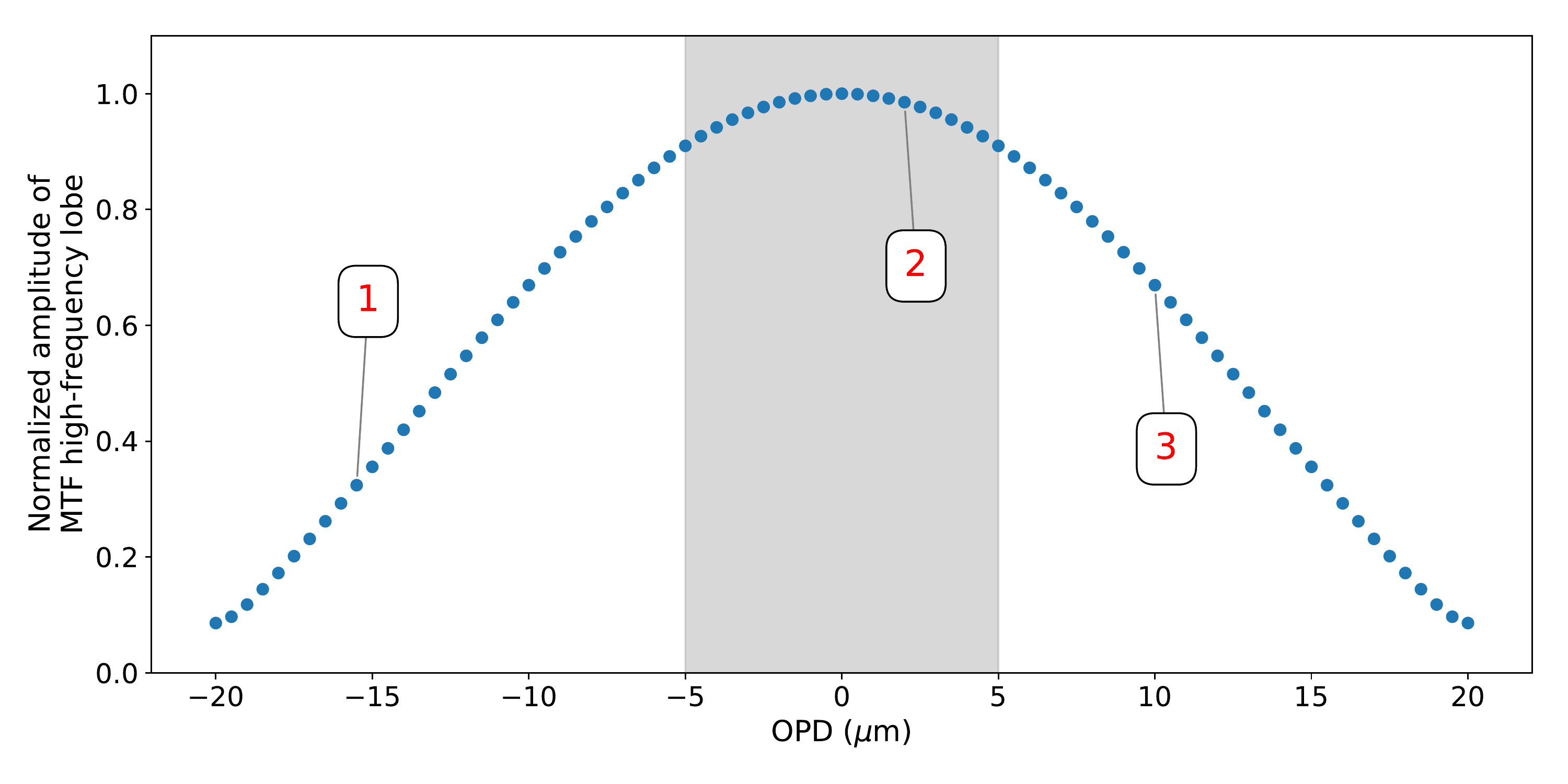}
   \end{tabular}
   \end{center}
   \vspace{-0.6cm}
   \caption[example] 
   { \label{fig:greystuff} 
Left: Fringe angles as found from empirical grism illuminations on LMIRcam, with a best-fit tangent line in grey. Angles $0\degree < \theta < 5\degree$ have been masked to avoid confusion with power in Fourier space from low frequencies. Numbered points correspond to the illuminations in Fig.\ \ref{fig:grism_demos}. Right: The amplitude of the high-frequency lobe in the MTF of simulated polychromatic 3.4-4.0 $\mu$m PSFs (corresponding to the `StdL' filter in Fig.\ \ref{fig:coh_size}), as a function of path length distance from the center of the coherence envelope. When the OPD is zero, the high-frequency fringes have maximum contrast. At nonzero OPD, the contrast washes out as different wavelengths are at different levels of constructive and destructive interference. The grey region spans a range of $\pm5$ $\mu$m, which is the allowable range of optical path change before the phase loop opens. Numbered points correspond to the illuminations in Fig.\ \ref{fig:airy_demos}.}
\end{figure}

The center of the coherence envelope can also be found without a grism by scanning in OPD, fitting a curve to the amplitude of the high-frequency lobe of the MTF, and shifting the HPC to put the optical path difference at the center of the coherence envelope (see Figs. \ref{fig:airy_demos}, \ref{fig:greystuff}).

\subsection{The correction code}
\label{subsec:corrxn_code}

\begin{figure} [ht]
   \begin{center}
   \begin{tabular}{c} 
   \includegraphics[width=0.47\linewidth, trim={5cm 5.3cm 5cm 8cm}, clip=True]{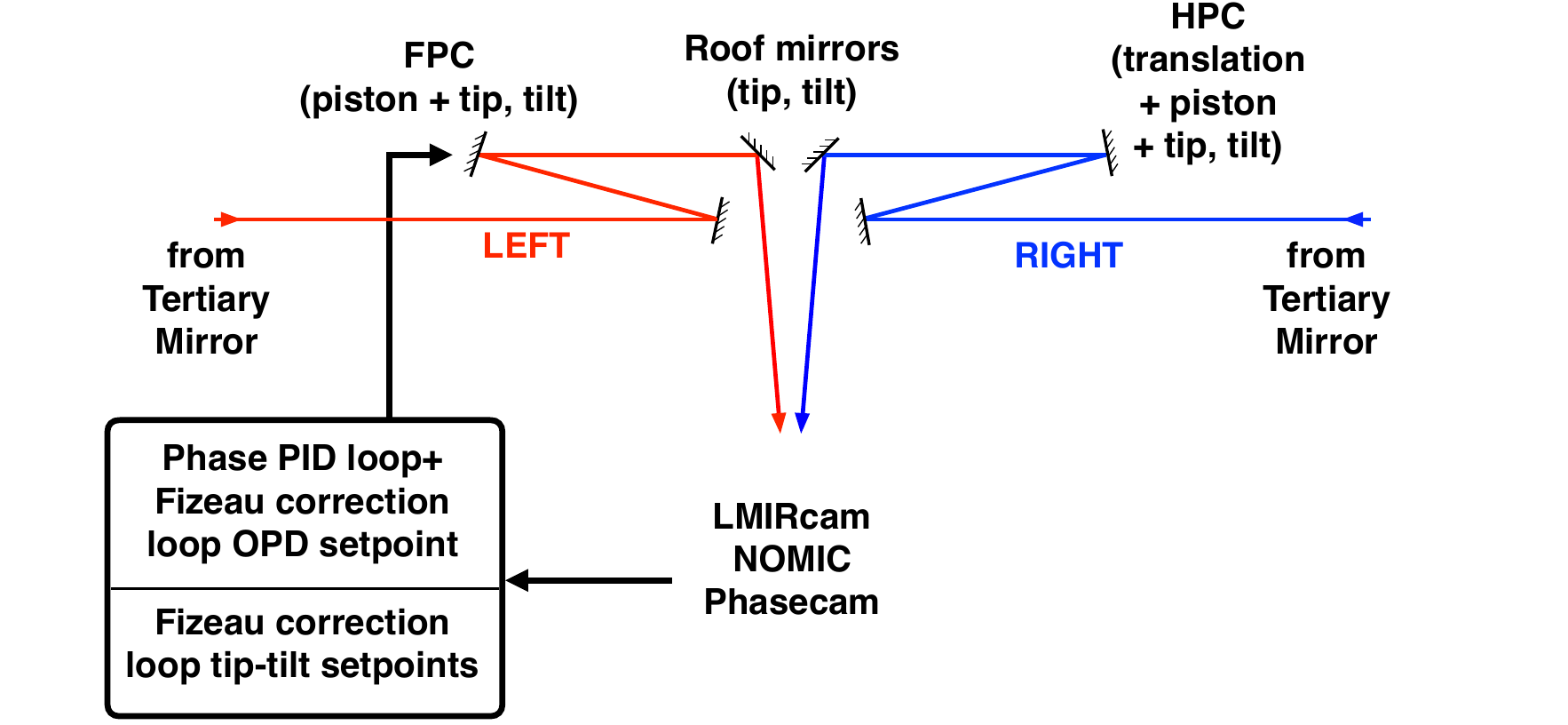}
   \hspace{0.5cm}
   \includegraphics[width=0.42\linewidth, trim={5cm 5cm 3cm 6cm}, clip=True]{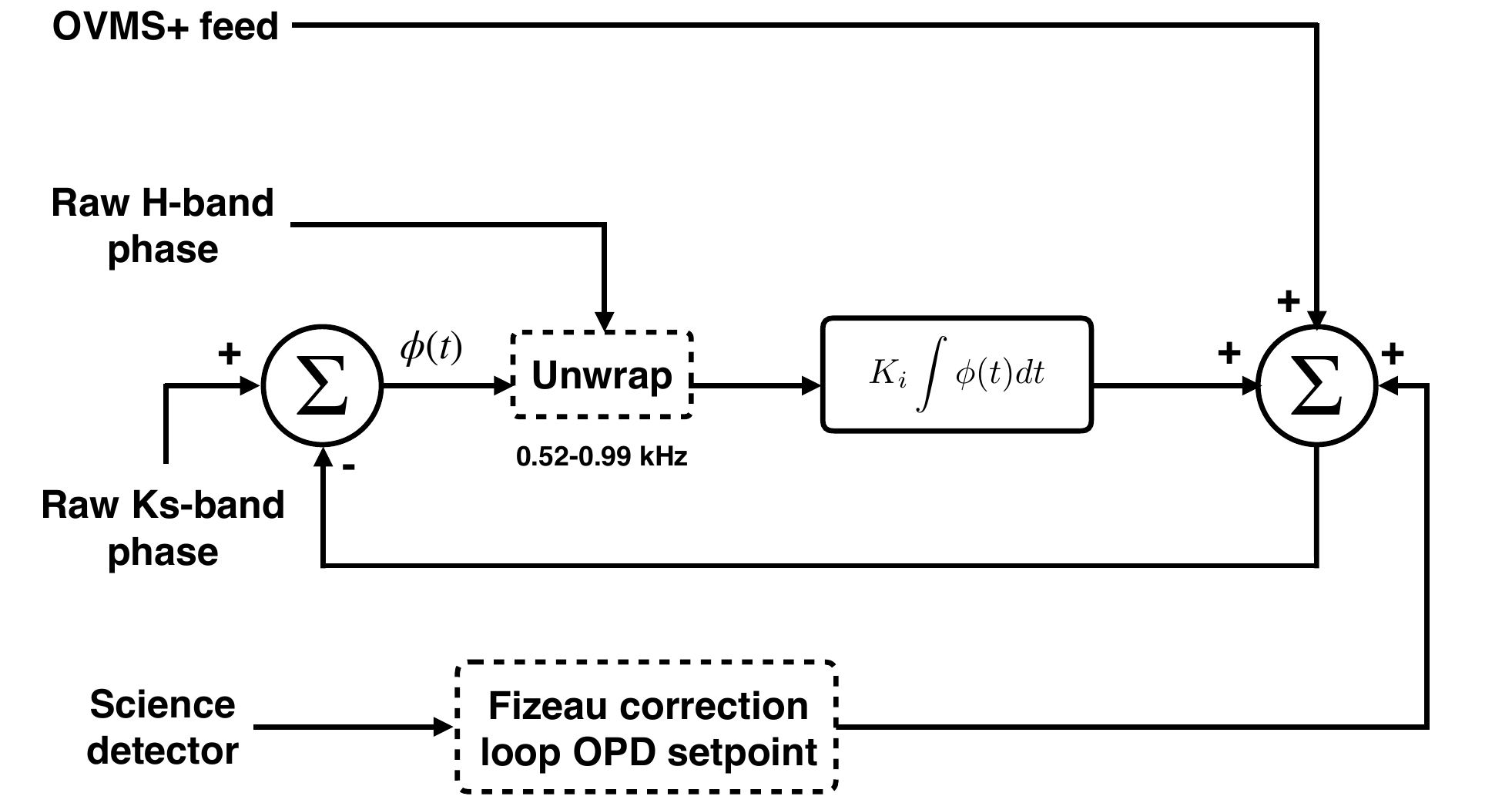}
   \end{tabular}
   \end{center}
   \vspace{-0.4cm}
   \caption[example] 
   { \label{fig:pid_diag} 
Left: The basic mechanism for removing NCPA with a Fizeau correction loop. Readouts from the science detectors in LMIRcam and/or NOMIC are analyzed and are used to calculate setpoints of the PhaseCam PID phase loop. Right: A schematic of the current phase correction, with dashed boxes to indicate components under development. Note the current ``PID'' loop currently just contains the ``I'' (integral-over-history) portion. (Compare with Fig.\ 5 in \cite{defrere2016nulling}.)}
\end{figure}

Once the OPD is small enough so that it is well within the coherence envelope of PhaseCam, and if line-of-sight seeing allows, the phase loop can be closed. This loop sends corrective movement commands to the FPC mirror (Fig.\ \ref{fig:pid_diag}).

But even with a closed phase loop, aberrations can appear on the science detectors. The correction code, which supplements the phase loop, uses readouts from the science detectors and 

\begin{enumerate}
    \item Finds the coarse centroid of the PSF by smoothing the detector subarray and finding the pixel with the maximum number of counts.
    \item Makes a cut-out of the subarray around the centroid.
    \item Fast-Fourier transforms the cut-out.
    \item Analyzes the MTF (amplitude) and PTF (phase) of the transform.
\end{enumerate}

Whereas the location within the coherence envelope can be sensed using the amplitude of the MTF, the differential aberrations described in Sec.  \ref{sec:intro} can be sensed using the slope of the PTF (for tip-tilt), or by detecting a stairstep pattern in the PTF (for OPD). 

Differential tip $\Theta_{y}$ and tilt $\Theta_{x}$ can be calculated from the slope of the PTF $\Omega_{y}$ in y and $\Omega_{x}$ in x as

\begin{equation}
\vec{\Theta}
=
\begin{bmatrix}
    \Theta_{x} \\
    \Theta_{y}
\end{bmatrix}
=
\begin{bmatrix}
    \Omega_{x}N_{x} \\
    \Omega_{y}N_{y}
\end{bmatrix}
\left(\frac{PS\cdot\Delta}{\pi} {\rm pix}_{DFT} \right)
\label{eqn:ptf}
\end{equation}

\noindent
where $N_{i}$ is the number of pixels along the $i$ axis of the subarray to be Fourier transformed, $PS$ is the plate scale\footnote{For LMIRcam, 10.7 mas/pix$_{det}$ \cite{spalding2019dewarp}; for NOMIC, 18 mas/pix$_{det}$ \cite{hoffmann2014operation}.}, and $\Delta$ is the sampling spacing in the plane of the detector (i.e., one detector pixel pix$_{det}$). The unit pix$_{DFT}$ is one  `pixel' in the discrete Fourier transform of the image. (See Appendix \ref{sec:appendix_ptf} for a derivation.)

\begin{figure} [!htb]
   \begin{center}
   \begin{tabular}{c} 
   \includegraphics[width=0.88\linewidth]{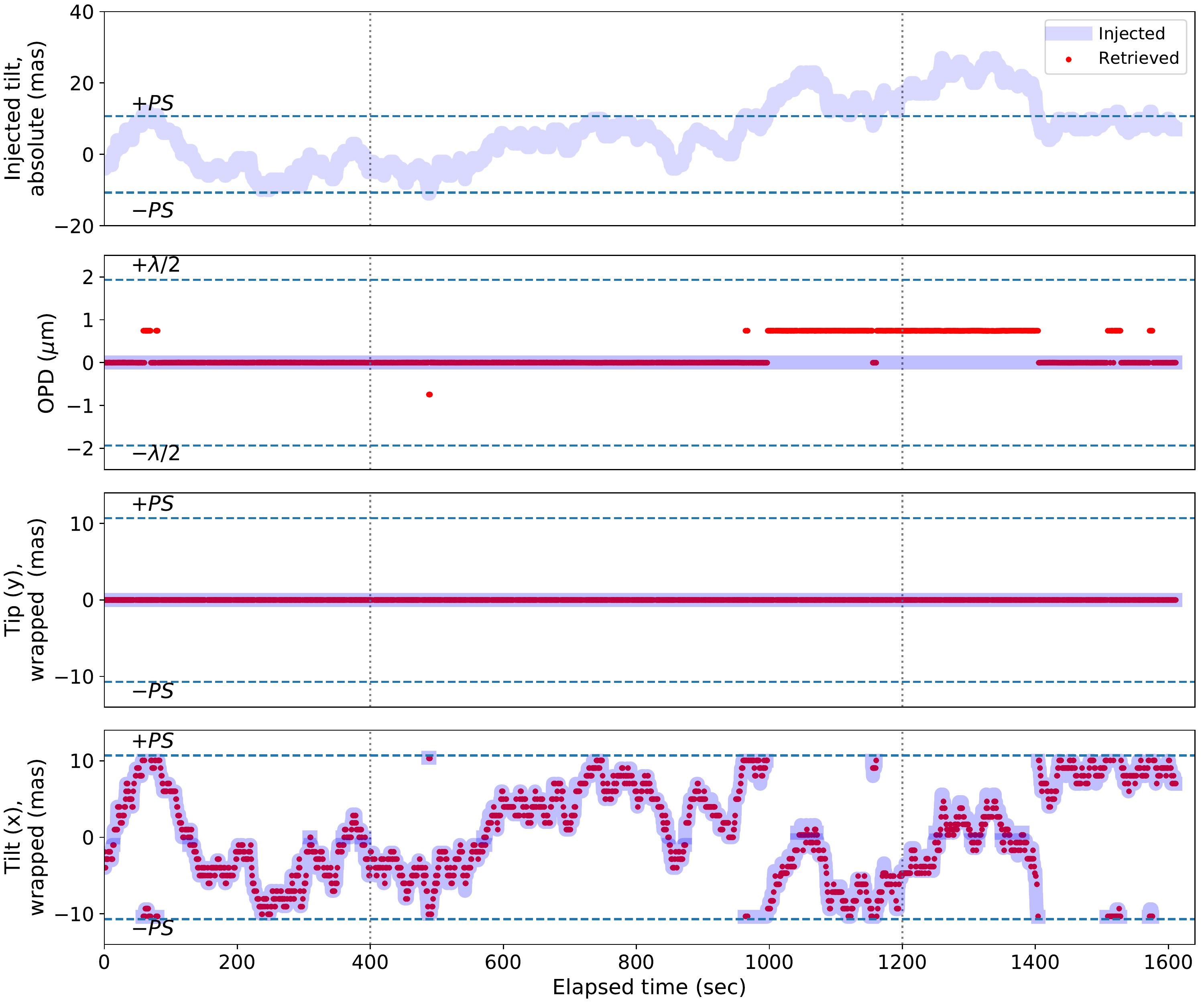} \\
    \hline \\
    \vspace{-0.8cm} \\
    \includegraphics[width=0.18\linewidth]{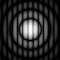}
    \includegraphics[width=0.33\linewidth, trim={0cm 1cm 0cm 0cm}, clip=True]{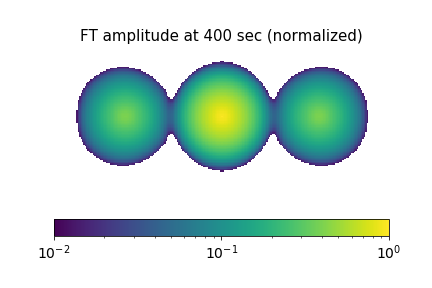}
    \includegraphics[width=0.33\linewidth, trim={0cm 1cm 0cm 0cm}, clip=True]{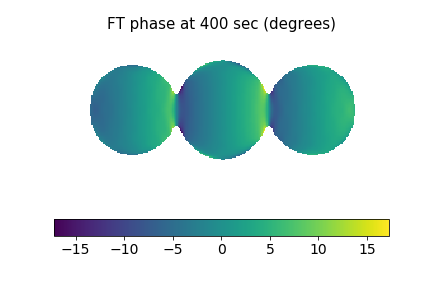} \\
    \vspace{-0.4cm} \\
     \hline \\
     \vspace{-0.8cm} \\
    \includegraphics[width=0.18\linewidth]{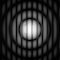}
    \includegraphics[width=0.33\linewidth, trim={0cm 1cm 0cm 0cm}, clip=True]{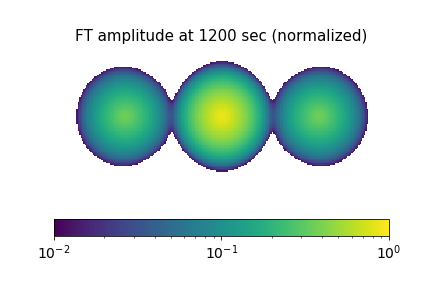}
    \includegraphics[width=0.33\linewidth, trim={0cm 1cm 0cm 0cm}, clip=True]{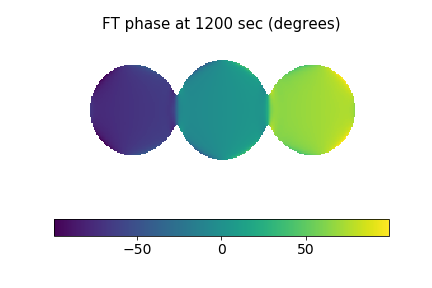} \\
    \vspace{-0.4cm} 
   \end{tabular}
   \end{center}
   \caption[example] 
   { \label{fig:tilt_spie} 
Simulated retrieval of differential tilt. Injected tilt takes a random walk, while the injected tip and OPD are zero. In the current version of the correction code, tilt can masquerade as OPD if the absolute tilt is greater than the angle corresponding to the plate scale (PS) for one pixel. Dotted lines indicate the locations in time of the PSFs (in logarithmic greyscale), MTFs, and PTFs shown in the bottom rows.}
\end{figure} 

\begin{figure} [ht]
   \begin{center}
   \begin{tabular}{c} 
   \includegraphics[width=0.9\linewidth]{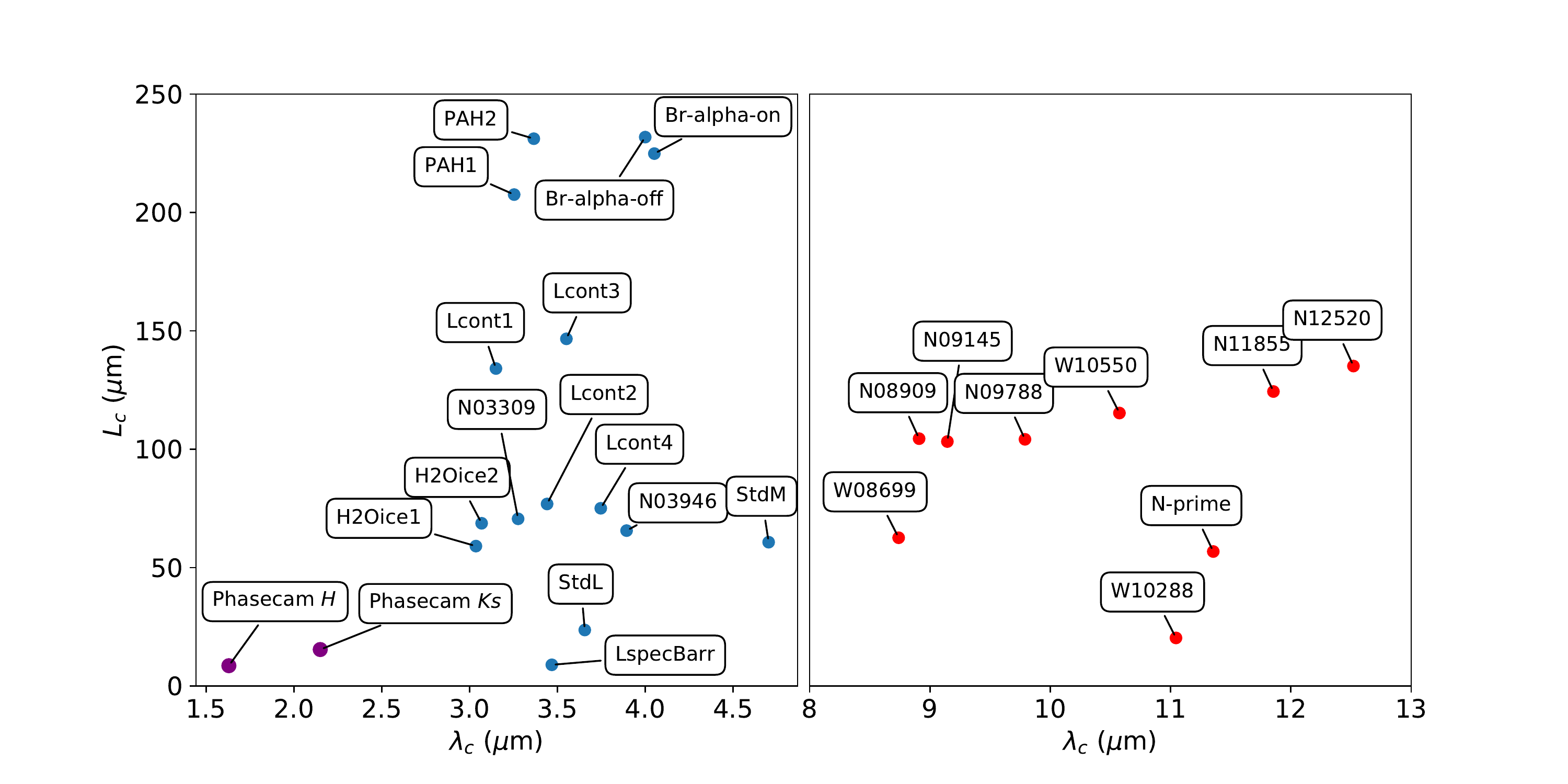}
   \end{tabular}
   \end{center}
   \vspace{-0.6cm}
   \caption[example] 
   { \label{fig:coh_size} 
Coherence envelope sizes $L_{c}=\lambda_{c}^{2}/\Delta \lambda$ of various filters for PhaseCam (purple), LMIRcam (blue), and NOMIC (red). Envelopes can be expanded at the science detectors with the use of grisms.}
\end{figure} 


\subsection{Code performance on simulated data}

Simulated data was generated with monochromatic, 3.7 $\mu$m, diffraction-limited LBT Fizeau PSFs with three degrees of freedom: OPD and differential tip and tilt. Three synthetic datasets of 10k frames were generated: one dataset in which the OPD did a random walk from frame to frame, while differential tip and tilt remained zero; a second dataset in which only tip did a random walk, and a third dataset in which only tilt did a random walk.

Fig.\ \ref{fig:tilt_spie} shows an example retrieval, using Eqn. \ref{eqn:ptf}, for a dataset in which the differential tilt does a random walk. The retrieved values wrap around the positive or negative plate scale, though in principle it is possible to break this degeneracy since the PSF will elongate as the Airy PSFs move apart.

\subsection{On-sky engineering tests}

Ultimately, tests of our correction code must be done on-sky: genuine Fizeau PSFs manifest the imperfect AO correction, NCPA effects, optical ghosts, speckles corresponding to the PSFs of the individual unit telescopes of the LBT, time-dependent detector and photon noise, and phase noise (especially if the phase loop is open, in which case there is significant phase ``smearing'', even in fast readouts).

In the fall of 2018 and spring of 2019 we carried out a series of on-sky engineering tests of different parts of the Fizeau correction code. Our objectives were to test routines in the consecutive order in which they would be executed for science observations: the overlapping of the Airy PSFs, putting in a grism and dialing the OPD to find the center of the coherence envelope, removing the grism, and then calculating PhaseCam setpoints in realtime as data is being taken. 

\subsection{Mechanical issues}

Once corrective movements are calculated---be they setpoints to the phase PID loop or explicit mirror movement commands---the quality of the implementation is an additional issue to consider. It should be noted that the fast and slow pathlength corrector mirrors are currently operating without direct feedback about where the mirrors actually are. \footnote{Capacitive sensors were originally built in to the design, but it was found that their response rates conflicted with the phase PID loop. In addition, at cryogenic temperatures the gap between the capacitive plates increases to the point where their sensitivity was lost, unless large voltages were applied, in which case there would be greater risk of shorting.} Setpoints to the phase PID loop are reliably implemented, because the phase PID loop seeks to match the setpoints with the Fourier-space image of the PhaseCam illumination. However, for alignment during setup or work in open-phase-loop, hysteresis in the mirror PZTs can be a problem. 

\begin{figure} [!ht]
   \begin{center}
   \begin{tabular}{c} 
   \includegraphics[height=7cm, trim={0cm, 0cm, 0cm, 0.9cm}, clip=True]{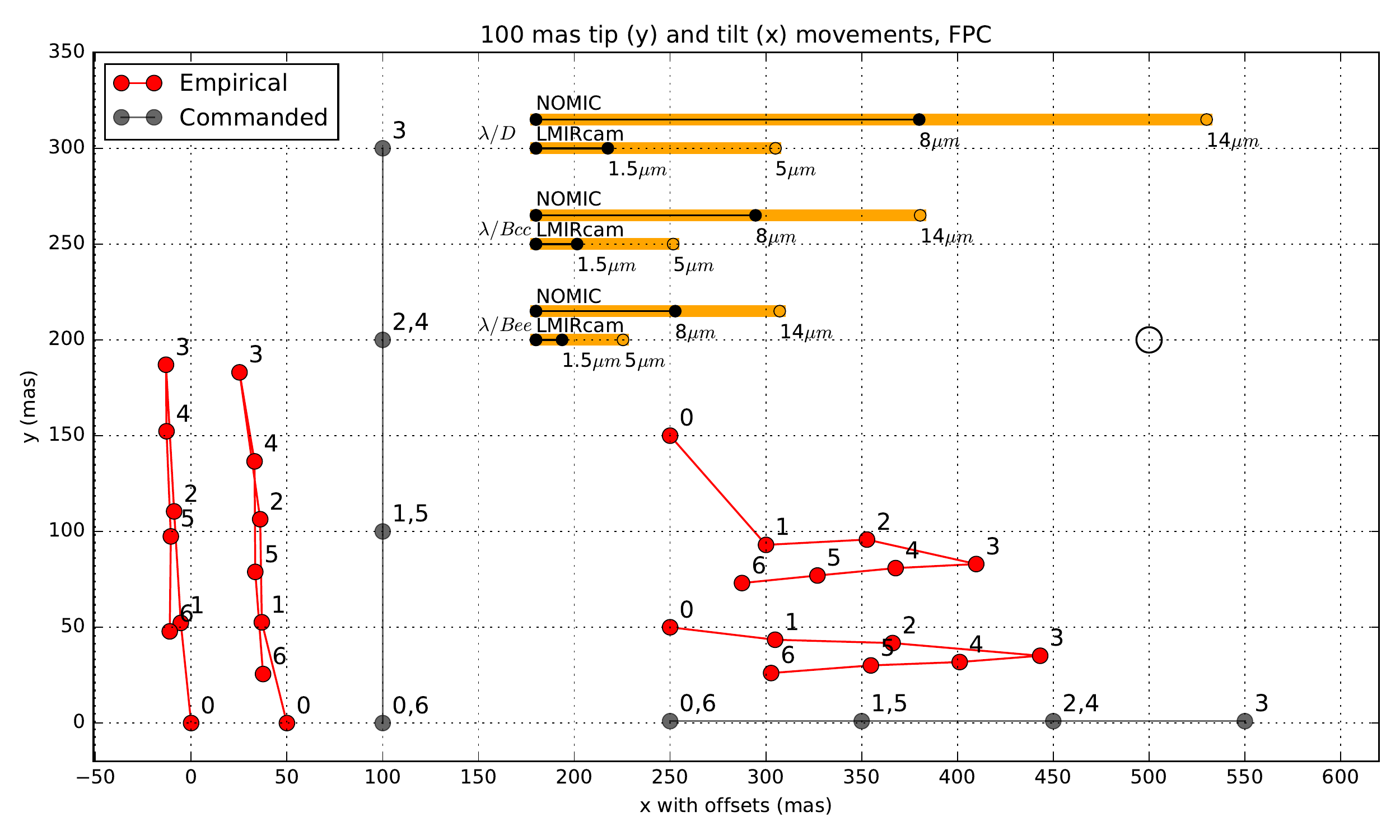}
   \end{tabular}
   \end{center}
   \vspace{-0.9cm}
   \caption[example] 
   { \label{fig:fpc_hyst_large} 
FPC hysteresis for 100 mas commanded movements. Commanded tip (y) and tilt (x) movements are in grey, with empirical results from two different trials for each in red. There were two sets of tip-tilt hysteresis tests, separated by thermal cycles and months of time. Numbers indicate the sequence of positions; i.e., 0 $->$ 1 $->$ 2 $->$ 3 etc. Hysteresis leads to the offsets between step pairs 0/6, 1/5, and 2/4. In terms of the offset distance between points 0 and 6 over the total movement, the hysteresis is $\approx10-20$ \%. (The undershooting by $\approx50$ is due to an underestimated scale factor in the software.) Orange and black bars in the upper right indicate relevant scales ($\lambda/D$ from one primary mirror, $\lambda/Bcc$ for the center-to-center baseline between the primary mirrors, and $\lambda/Bee$ for the edge-to-edge baseline between the primary mirrors) at wavelengths accessible to the two cameras. To provide a rough sense of the centroiding uncertainty, the unfilled circle has a radius of the mean radial distance of a set of eight found PSF positions from the average position, during a sequence of frames where the FPC was not commanded to move.}
\end{figure}

We tested for hysteresis in the FPC and HPC mirrors by finding the centers of thermal pinhole images of the telescope chamber. The pinhole locations were found with the \texttt{astropy} implementation of \texttt{DAOPHOT} \cite{stetson1987daophot,price2018astropy}. In Fig.\ \ref{fig:fpc_hyst_large} we show an example of a hysteresis test on the FPC. We find that, for commanded movements as large as 100 mas over a total commanded range of 600 mas, hysteresis leads to $\approx10-20$ \% positional uncertainty. This repeatability allows for counteraction by re-scaling the commanded mirror movements.

\section{Lessons learned from on-sky tests}
\label{sec:lessons}

Some of the lessons we learned from on-sky tests in the fall of 2018 and spring of 2019 were as follows:
\begin{itemize}
\item When iteratively moving the unit telescopes themselves (and not instrument mirrors) to overlap the Airy PSFs, there can be slop if the commanded movement is equivalent to the distance on the detector from the current to the desired pixel location and is $\gtrsim$1 asec.  Convergence may be made more efficient by rescaling the commanded movements. 
\item If the Airy illuminations have been overlapped on the science detector, inserting a grism upstream introduces a focus offset and causes the grism illuminations to spring apart on the detector by a fraction of an arcsecond. This small separation is tolerable insofar as the fringes are still distinct enough for bringing the OPD to zero, and the additional overhead of overlapping the grism illuminations is not necessary, unless the science observation itself will be in Fizeau-grism mode. (See Sec.  \ref{subsec:initial_align}).  
\item The phase loop can be closed after the AO SOUL upgrade. On UT 2019 Feb 24 the PhaseCam loop was closed for the first time following the SOUL upgrade on both telescopes, for up to roughly half a minute at a time in good but somewhat unstable seeing. For now, however, it would be advisable to keep the AO correction at 1 kHz for interferometry because SOUL has been found to be rather unstable in tip-tilt at faster speeds, which can break the phase loop. 
\item Side-to-side nodding with the telescopes in Fizeau mode is repeatable. This nodding is done by physically moving the telescopes, and also a lens wheel upstream of PhaseCam which has pairs of lenses at staggered radial positions. Switching from one set of lenses to another re-centers the illumination on PhaseCam after the telescopes have been moved. 
\item In open-phase-loop Fizeau, it might be preferable to nod up-down with the telescopes to avoid introducing OPD changes. 
\item If fringes have high visibility on the science detector in Fizeau-grism mode, we can retrieve OPD values that reliably correct the gross path length using the HPC mirror. This is the case even when the phase loop is open and there is atmospheric jitter in the fringes. 
\item If PhaseCam pupils are well-aligned and the phase loop is closed, but the PSFs are not aligned on LMIRcam, one can adjust a pupil steering mirror (PSM) upstream of PhaseCam in small amounts. The phase PID loop will keep the fringes aligned as before on PhaseCam, but the alignment on LMIRcam will change because the PID loop will move the FPC. But this is also not time-efficient, so it is best to complete the co-alignment in open phase loop if possible.
\item There is a risk of translation stages getting stuck after a cryo-cycle. One (unproven) possibility is that volatiles migrate to the bearings when warming up the cryostat. It is adviseable to exercise the stages before cool-down, and before interferometry.
\end{itemize}

\section{Conclusion and Future directions}
\label{sec:future}

The software- and hardware-based commissioning of LBTI's Fizeau mode is underway. Over the past year, we have taken off-sky and on-sky engineering data for the purposes of writing code to automatize the alignment process, and remove NCPA during observations. There is still development and a number of improvements yet to be made, however, and we have been granted more on-sky engineering time in the fall 2019 observing season to do so. 

In the meantime, we are working on data reduction pipelines for Fizeau data which has already been taken, and we continue to take Fizeau observations among the various science programs of the LBTI queue. In spring 2018, there were three Fizeau science programs (that is, not engineering programs) in the LBTI queue; in fall 2018, four; and in spring 2019, two.

In fall 2019 we will also begin installing  capacitive sensors behind the corrector mirrors to provide closed-loop feedback on the mirror positions. We will start by upgrading the HPC, and, if this is successful, we will upgrade the FPC.

Further into the future, all of the current Fizeau correction code will fade from Python into a lower-level language like C. We will also build a webpage interface for the Fizeau mode, modeled on our PhaseCam control webpage, so as to make Fizeau mode observations as user-friendly and as useful as possible.

\appendix

\section{Glossary}
\label{sec:appendix_glossary}

\begin{itemize}
\item AO: adaptive optics
\item FPC: fast pathlength corrector; this mirror can be used to adjust tip, tilt, and small amounts of pathlength at up to 1 kHz (Fig.\ \ref{fig:pid_diag})
\item HOSTS: LBTI survey of exozodiacal dust disks within 30 parsecs \cite{ertel2018hosts}
\item HPC: hybrid pathlength corrector, consisting of a slow pathlength corrector mounted on a translation stage for large pathlength adjustments (Fig.\ \ref{fig:pid_diag})
\item INDI: telescope and instrument control software
\item LMIRcam: 1.2--5 $\mu$m science camera
\item LBT(I): Large Binocular Telescope (Interferometer)
\item MTF: modulation transfer function; the amplitude of the OTF 
\item NCPA: non-common-path aberrations
\item NOMIC: 8--12 $\mu$m science camera 
\item NRM: non-redundant phase masking
\item OPD: optical path difference 
\item OTF: complex optical transfer function, describing the amount of spatial information transfer; this can be calculated as the Fourier transform of the PSF: $FT\left\{PSF(x,y)\right\} \equiv  OTF(\zeta,\eta) \equiv MTF(\zeta,\eta)exp[-iPTF(\zeta,\eta)]$
\item OVMS+: the telescope vibration monitoring and calculated compensation system \cite{bohm2016ovms}
\item PhaseCam: $H$- and $Ks$-band phase camera  
\item PID: proportional-integral-differential control loop (specifically implemented to use phase information to control the FPC mirror)
\item PSM: pupil steering mirror, upstream of PhaseCam and NOMIC 
\item PTF: phase transfer function; the phase of the OTF
\item PZT: ceramic material which adjusts in length depending on applied voltage
\item SOUL: the upgrade to the LBT adaptive optics systems, implemented in the fall of 2018 and early 2019 \cite{pinna2016soul}
\end{itemize}

\newpage
\section{Alignment as seen by the cameras}
\label{sec:appendix_align}

This section illustrates the various illuminations on the camera detectors in the course of making the optical alignments for phase-controlled Fizeau interferometry with LBTI.

\begin{figure} [ht]
   \begin{center}
   \begin{tabular}{l} 
   \includegraphics[width=1\linewidth]{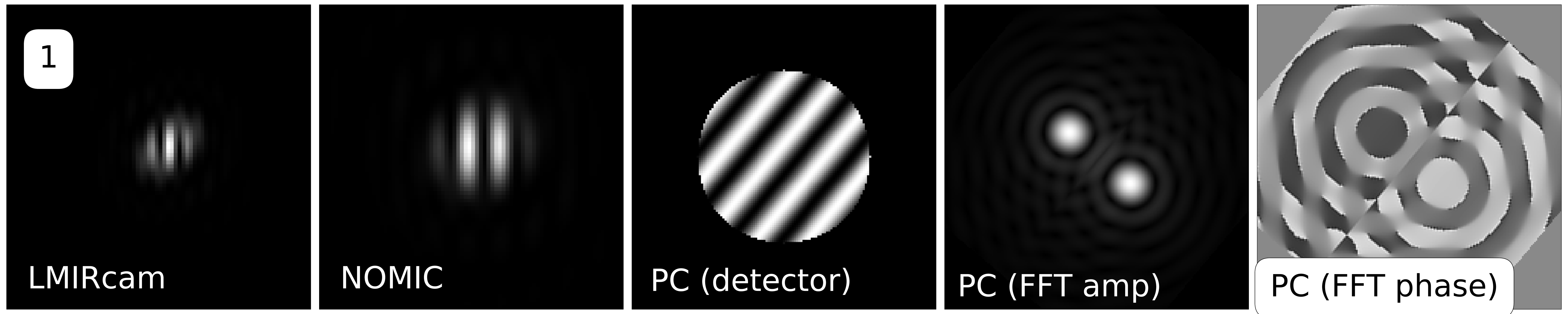} \\
   \begin{small}
   \begin{tabular}{@{}l@{}l@{}}\textbf{Just before closing phase loop.}\\Science detector Airy PSFs have been roughly overlapped, but\\the Fizeau PSFs exhibit OPD, tip, and tilt aberrations.\\Fringes are visible on all detectors, but are moving between \\frames. Fringes on PhaseCam (PC) are at a random angle.\end{tabular}
   \hspace{2cm}
   \begin{tabular}{@{}l@{}l@{}}Pathlength setpoint: 0\\Tip setpoint: 0\\Tilt setpoint: 0\end{tabular}
    \end{small} \\
   \hline \\
   \includegraphics[width=1\linewidth]{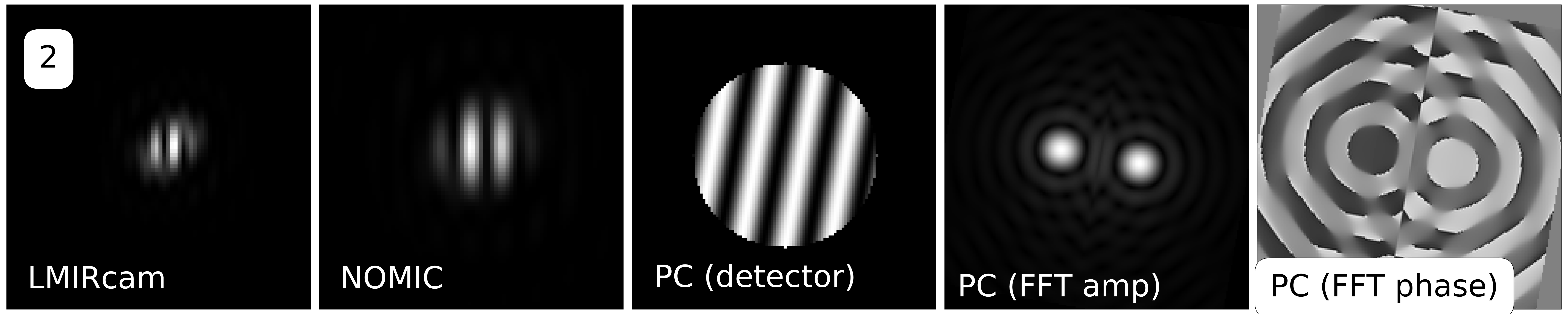} \\
   \begin{small}
   \begin{tabular}{@{}l@{}l@{}}\textbf{Just after closing phase loop.}\\PhaseCam PID phase loop analyzes the Fourier transforms of the\\PhaseCam image and starts to adjust FPC in piston to put\\bright fringe at center, and FPC in tip/tilt to rotate the fringes vertically\end{tabular}
   \hspace{0.4cm}
   \begin{tabular}{@{}l@{}l@{}}Pathlength setpoint: 0\\Tip setpoint: 0\\Tilt setpoint: Nominal nonzero value.\end{tabular}
    \end{small} \\
   \hline \\
   \includegraphics[width=\linewidth]{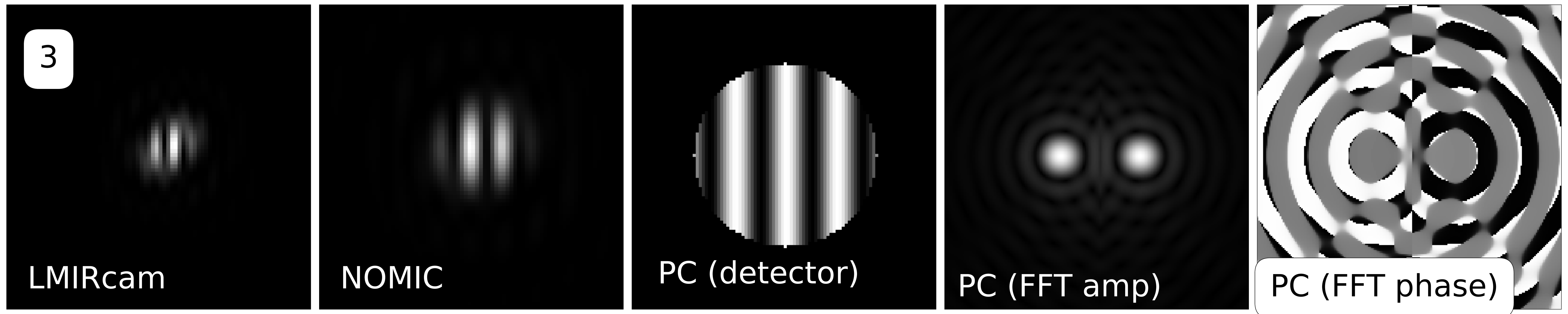} \\
   \begin{small}
   \begin{tabular}{@{}l@{}l@{}}\textbf{Phase loop has converged.}\\A bright fringe is at the center of the PhaseCam pupil, though\\the OPD is not necessarily at the center of the coherence envelope\\on PhaseCam. 
    \end{tabular}
   \hspace{1.5cm}
   \begin{tabular}{@{}l@{}l@{}}Pathlength setpoint: 0\\Tip setpoint: 0\\Tilt setpoint: Nominal nonzero value.\end{tabular}
    \end{small} \\
    \hline
   \end{tabular}
   \end{center}
   \vspace{-0.4cm}
   \caption[example] 
   { \label{fig:cartoon_1} 
A simulated sequence of images to represent what one would see with the science detectors and PhaseCam during the alignment sequence, and after perturbations to the phase loop setpoints. (Panels continue in Fig.\ \ref{fig:cartoon_2}.)}
\end{figure}

\begin{figure} [ht]
   \begin{center}
   \begin{tabular}{l} 
   \hline
   \includegraphics[width=\linewidth]{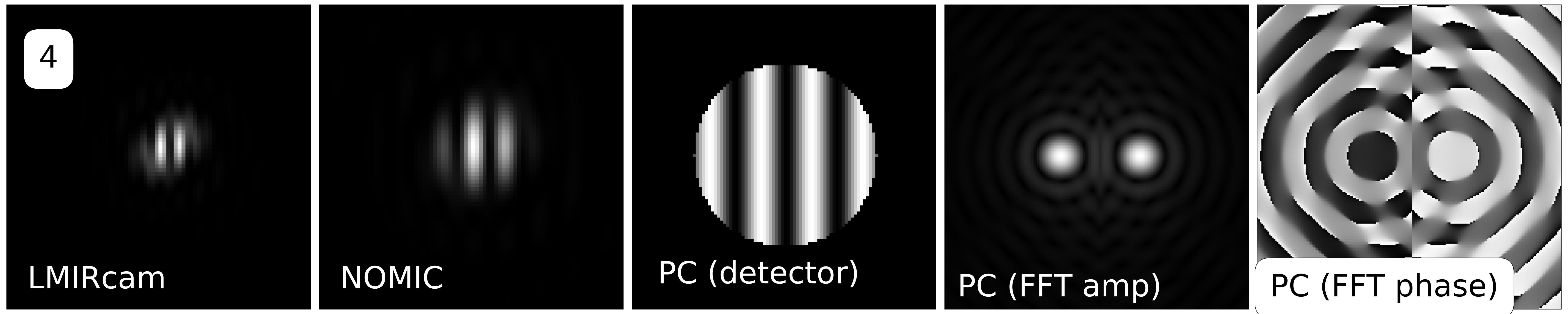} \\
   \begin{small}
   \begin{tabular}{@{}l@{}l@{}}\textbf{Nonzero pathlength setpoint has been sent.}\\Fringes on the PhaseCam pupil are now offset. OPD aberrations\\are now zero on the science detectors.\\
    \end{tabular}
   \hspace{1.7cm}
   \begin{tabular}{@{}l@{}l@{}}Pathlength setpoint: 180 degrees\\Tip setpoint: 0\\Tilt setpoint: Nominal nonzero value.\end{tabular}
    \end{small} \\
   \hline
   \includegraphics[width=1\linewidth]{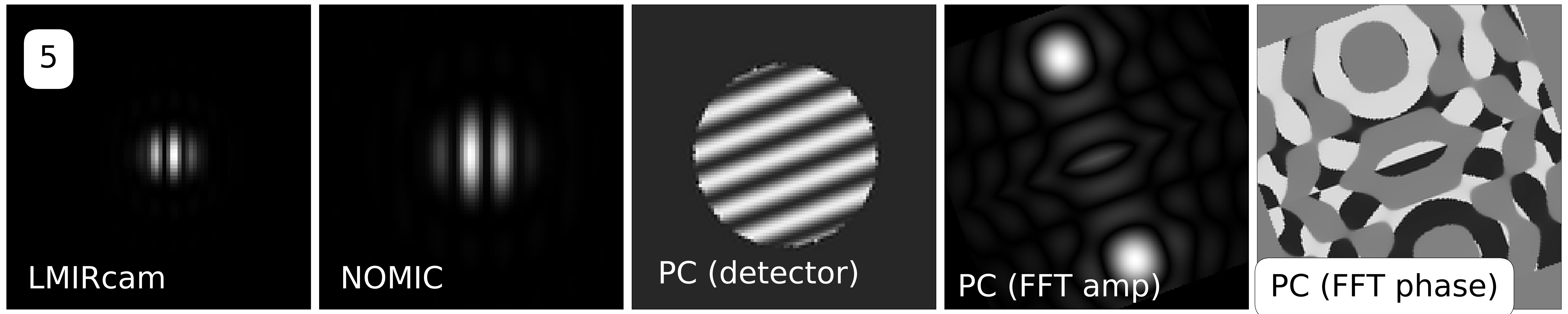} \\
   \begin{small}
   \begin{tabular}{@{}l@{}l@{}}\textbf{Nonzero tip setpoint has been sent.}\\Fringes on the PhaseCam pupil are rotated because both tip (y) and\\ tilt (x) setpoints are nonzero. Tip aberrations are now zero on the\\science detectors.\end{tabular}
   \hspace{1.1cm}
   \begin{tabular}{@{}l@{}l@{}}Pathlength setpoint: 0\\Tip setpoint: Nonzero\\Tilt setpoint: Nominal nonzero value.\end{tabular}
    \end{small} \\
   \hline \\
   \includegraphics[width=1\linewidth]{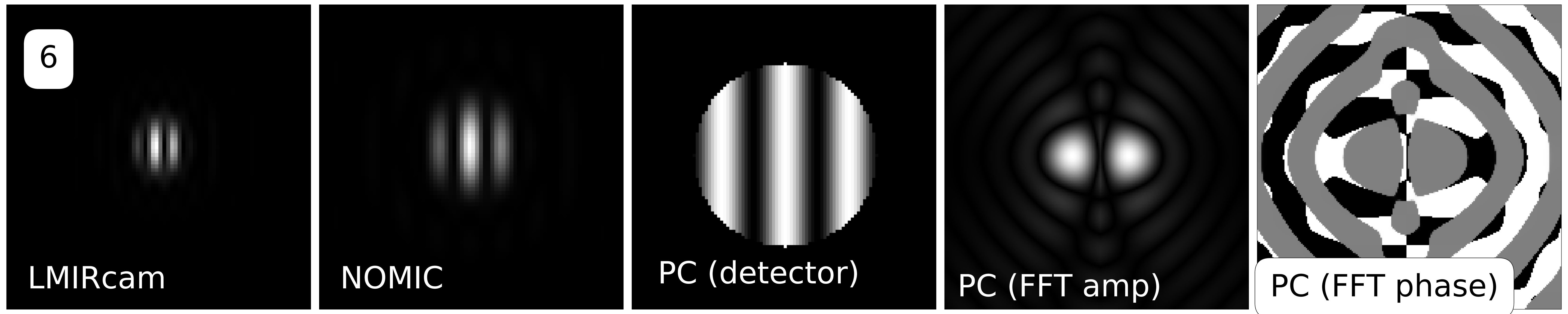} \\
   \begin{small}
   \begin{tabular}{@{}l@{}l@{}}\textbf{Nonzero, non-nominal tilt setpoint has been sent.}\\Fringes on the PhaseCam pupil assume a different frequency\\along the horizontal. Tilt aberrations are now zero on the\\science detectors.\end{tabular}
   \hspace{2.1cm}
   \begin{tabular}{@{}l@{}l@{}}Pathlength setpoint: 0\\Tip setpoint: 0\\Tilt setpoint: Nonzero, non-nominal value.\end{tabular}
    \end{small} \\
   \hline \\
   \includegraphics[width=1\linewidth]{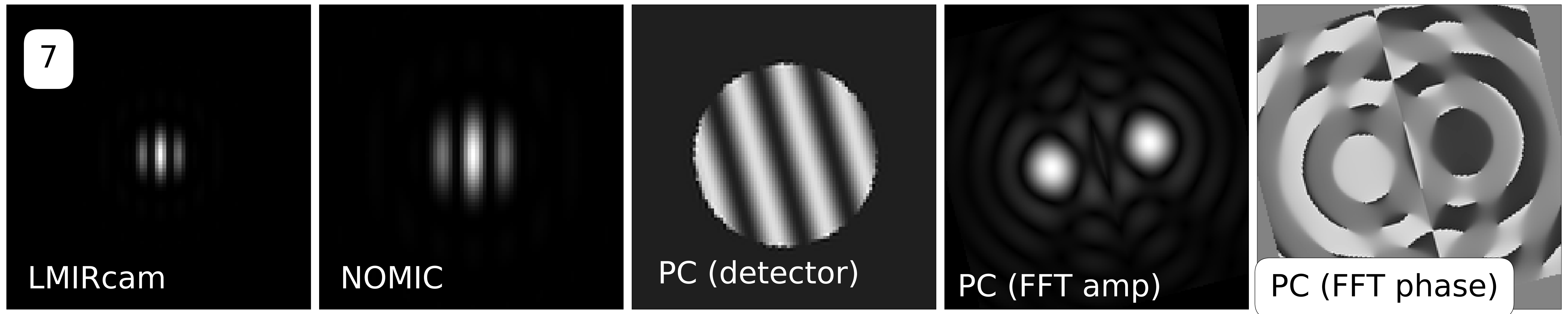} \\
   \begin{small}
   \begin{tabular}{@{}l@{}l@{}}\textbf{Based on the Fourier transforms of the science detector}\\\textbf{illuminations, setpoints for OPD, tip, and tilt have been}\\\textbf{sent to remove all three of these aberrations from}\\\textbf{the science PSFs.}\\ Fringes on the PhaseCam pupil assume a frequency and rotation\\corresponding to these setpoints.\end{tabular}
   \hspace{1.5cm}
   \begin{tabular}{@{}l@{}l@{}}Pathlength setpoint: Nonzero\\Tip setpoint: Nonzero\\Tilt setpoint: Nonzero, non-nominal value.\end{tabular}
    \end{small} \\

    \hline
   \end{tabular}
   \end{center}
   \vspace{-0.4cm}
   \caption[example] 
   { \label{fig:cartoon_2} 
Continuation of Fig.\ \ref{fig:cartoon_1}.}
\end{figure} 

\section{The PTF slope}
\label{sec:appendix_ptf}

Here we derive the wavefront tip/tilt from the slope of the PTF. Symbols are tabulated below.

\begin{table}[H]
\begin{center}
\caption{PTF Quantities} 
\label{table:ptf_quantities}
\vspace{-0.3cm}
\begin{tabular}{| c | l | l |}
\hline
\textit{Symbol}	& \textit{Quantity} & 	\textit{Units} \\
 \hline	
$N$	& \begin{tabular}{@{}l@{}}Number of detector pixels along the edge of a detector subarray\\which is to be Fourier transformed\end{tabular} &--\\
 \hline	
$\Lambda$	& Linear shift on science detector & pix$_{det}$\\
\hline
$\Omega$ & PTF slope  	& radians$\cdot$pix$_{DFT}^{-1}$ \\	
\hline
$PS$ & Detector plate scale  	& arcsec$\cdot$pix$_{det}^{-1}$ \\	
\hline
$\Theta$ & Wavefront tip/tilt with the normal  	& arcsec \\	
\hline
$\Delta$ & Sampling spacing in the plane of the detector & pix$_{det}$ \\
\hline
$\nu$, $\lambda$ & \begin{tabular}{@{}c@{}}Frequencies and wavelengths of information on the science detector\end{tabular} & pix$_{det}^{-1}$, pix$_{det}$ \\
\hline
\end{tabular}
\noindent 
\end{center}
\end{table}

At any given coordinate in Fourier space, the amplitude of the PTF represents the magnitude of the shift (in radians) of the wavelength corresponding to that Fourier coordinate.\footnote{Here, `wavelengths' refer to those bundled up in the detail in the image plane, and is not the science observing wavelength.} This implies that the PTF should be sloped for a pure translation of an image: different levels of detail have to be shifted by different numbers of constituent wavelengths to move in unison on the detector \cite{williams1989introduction}.

Consider a pure translation of an Airy function along the axis $+x_{det}$ on the detector, corresponding to a wavefront tilt of $\Theta_{x}$ relative to the normal. The distance moved on the detector in pixels is $\Lambda_{x}=\Theta_{x}/PS$, where $PS$ is the detector plate scale.

Without any loss of generality, we will drop the $x$ subscripts for now and treat this as a one-dimensional problem. The `pixel' coordinate in the Fourier transform which corresponds to the lowest frequency---other than the zero frequency, which is just the average illumination---corresponds to the frequency $\nu_{1}=1/N\Delta$, or the wavelength $\lambda_{1}=N\Delta$. Here, $N$ is the number of samples along $x_{det}$ across the image that will be Fourier transformed. The $\Delta$ is the sampling spacing (i.e., pix$_{det}$).

Now, consider a position in discrete Fourier space $k$ pix$_{DFT}$ higher than the frequency $\nu_{1}$ in the Fourier image. That pixel represents the frequency $\nu_{k}=k/(N\Delta)$, or the wavelength $\lambda_{k}=N\Delta/k$. For the original angular shift of $\Theta$, what is this shift in terms of radians of $\lambda_{1}$ and $\lambda_{k}$? In radians of $\lambda_{1}$, the movement on the detector is $2\pi\Lambda/\lambda_{1}=2\pi\Lambda/(N\Delta)$. For $\lambda_{k}$, it is $2\pi\Lambda/\lambda_{k}=2\pi k\Lambda/(N\Delta)$.

Generalizing, a straight-line PTF slope between any two Fourier pixels $i$ and $j$ is

\begin{equation}
\Omega = \frac{\Delta PTF}{\Delta x_{DFT}}\\
=\frac{(2\pi\Lambda/\lambda_{j})-(2\pi\Lambda/\lambda_{i})}{Dij}
\end{equation}

\noindent
where we use $D_{ij}$ to be the distance in pix$_{DFT}$ between $i$ and $j$---this avoids confusion with the other $\Delta$ floating around. Rewrite the wavelengths in terms of $N\Delta$ with scaling factors, viz. $\lambda_{i}=N\Delta/I_{i}$ and $\lambda_{j}=N\Delta/I_{j}$. Continuing our equality, we have \\

\begin{equation}
=\frac{2\pi\Lambda\left[\frac{I_{j}}{N\Delta}-\frac{I_{i}}{N\Delta}\right]}{Dij} \\
=\frac{2\pi\Lambda}{N\Delta}\frac{(I_{j}-I_{i})}{Dij}
\end{equation}

\noindent
Substituting in $\Lambda=\Theta/PS$, \\

\begin{equation}
=\frac{2\pi\Theta}{PS\cdot N\cdot\Delta}\frac{(I_{j}-I_{i})}{Dij}
\label{eqn:thiseqn}
\end{equation}

\noindent
Between Fourier pixels with indices $i=1$ and $j=k$, $I_{j}-I_{i}=k-1$, and the distance $D_{ij}$ between their centers is $(k-1)$ pix$_{DFT}$. Thus between any Fourier pixels $i$ and $j$, numbers cancel such that we are left with units alone in the rightmost piece of Eqn. \ref{eqn:thiseqn}: $\frac{(I_{j}-I_{i})}{Dij}={\rm pix}_{DFT}^{-1}$. Then we have \\

\begin{equation}
= \frac{2\pi\Theta}{PS\cdot N\cdot\Delta} \cdot {\rm pix}_{DFT}^{-1}
\end{equation}

\noindent
This is valid for a pure translation of an image an angular distance $\Theta$ with structure at different wavelengths. Indeed, this is just a manifestation of the translation property of Fourier transforms: a translation of an image by $\Theta$ in $x$ before a Fourier transform into $u$-space leads to an extra $2\pi\Theta u$ term in the phase of the Fourier transform.

But in Fizeau mode, when the left-side telescope Airy pattern is shifted with (for example) some tilt on the FPC mirror, the right-side Airy pattern stays put. The center of the net illumination on the detector shifts \textit{half} the distance of the the left-side Airy pattern. (Note that the illumination pattern \textit{around} its center is also changing, which causes the MTF to change.)

Thus we have translated the center of a modified image half the distance of the left-side Airy pattern, and $\Omega$ needs to be decreased by a factor of 2:  \\

\begin{equation}
\Omega = \frac{\pi\Theta}{PS\cdot N\cdot\Delta} \cdot {\rm pix}_{DFT}^{-1}
\end{equation}

\noindent
Generalizing to two dimensions, the PTF slope is 

\begin{equation}
\vec{\Omega}
=
\begin{bmatrix}
    \Omega_{x} \\
    \Omega_{y}
\end{bmatrix}
=
\begin{bmatrix}
    \Theta_{x}/N_{x} \\
    \Theta_{y}/N_{y}
\end{bmatrix}
\left(\frac{\pi}{PS\cdot\Delta} {\rm pix}_{DFT}^{-1} \right)
\end{equation}

\noindent
Rearranging to get the tip/tilt of the incident wavefront,

\begin{equation}
\vec{\Theta}
=
\begin{bmatrix}
    \Theta_{x} \\
    \Theta_{y}
\end{bmatrix}
=
\begin{bmatrix}
    \Omega_{x}N_{x} \\
    \Omega_{y}N_{y}
\end{bmatrix}
\left(\frac{PS\cdot\Delta}{\pi} {\rm pix}_{DFT} \right)
\end{equation}

\noindent
The required tip-tilt correction is then $\vec{\Gamma}=-\vec{\Theta}$. Note that these relations are independent of the science wavelength.  \\

\acknowledgments 

The LBT is an international collaboration among institutions in the United States, Italy and Germany. LBT Corporation partners are: The University of Arizona on behalf of the Arizona university system; Istituto Nazionale di Astrofisica, Italy; LBT Beteiligungsgesellschaft, Germany, representing the Max-Planck Society, the Astrophysical Institute Potsdam, and Heidelberg University; The Ohio State University, and The Research Corporation, on behalf of The University of Notre Dame, University of Minnesota and University of Virginia.

KMM's work is supported by the NASA Exoplanets Research Program (XRP) by cooperative agreement NNX16AD44G.

This research has made use of the Jean-Marie Mariotti Center JSDC catalogue\footnote{Available at \url{http://www.jmmc.fr/catalogue_jsdc.htm}}.

\bibliography{report} 
\bibliographystyle{spiebib} 

\end{document}